\documentclass[fleqn]{2023SCGE}
\usepackage{amsmath}
\usepackage{aas_macros}
\usepackage{booktabs}
\usepackage{array}
\usepackage{footnote}
\usepackage{listings}
\usepackage{amsmath}
\usepackage{multirow}
\usepackage{graphicx}
\usepackage{longtable}
\usepackage{pifont}
\usepackage{rotating}
\usepackage{tabularx}
\usepackage{ulem}
\usepackage{amssymb} 
\usepackage{mathrsfs}
\usepackage{adjustbox}
\usepackage{lscape}   
\usepackage{chngcntr}
\usepackage{supertabular,booktabs,array}
\begin{document}

\ensubject{subject}
\ArticleType{Article}

\title{Machine Phenomenology: A Simple Equation Classifying Fast Radio Bursts}{Machine Phenomenology: A Simple Equation Classifying Fast Radio Bursts}


\author[1,2,3]{Yang Liu}{}
\author[1,3]{Yuhao Lu}{}
\author[4]{Rahim Moradi}{}
\author[5]{Bo Yang}{}
\author[6,7,8]{Bing Zhang}{{bzhang1@hku.hk}}
\author[1, 5, 9, 12]{Wenbin Lin}{{lwb@usc.edu.cn}}
\author[3, 9, 10, 11]{Yu Wang}{{yu.wang@icranet.org}}

\AuthorMark{Y. Liu}

\AuthorCitation{Y. Liu, Y. Lu, R. Moradi, B. Yang, B. Zhang, W. Lin,  and Y. Wang}

\address[1]{School of Computer Science, University of South China, Hengyang 421001, China}
\address[2]{Department of Physics E. Pancini, University Federico II, Naples 80126, Italy}
\address[3]{ICRANet-AI, Brickell Avenue 701, Miami, FL 33131, USA}
\address[4]{Key Laboratory of Particle Astrophysics, Institute of High Energy Physics, Chinese Academy of Sciences, Beijing 100049, China}
\address[5]{School of Mathematics and Physics, University of South China, Hengyang 421001, China}
\address[6]{The Hong kong Institute for Astronomy and Astrophysics, University of Hong Kong, Pokfulam Road, Hong Kong, China}
\address[7]{Department of Physics, University of Hong Kong, Pokfulam Road, Hong Kong, China}
\address[8]{Nevada Center for Astrophysics, University of Nevada, NV 89154, USA}
\address[9]{ICRANet, P.zza della Repubblica 10, Pescara 65122, Italy}
\address[10]{ICRA and Dipartimento di Fisica, Sapienza Universit\`a di Roma, P.le Aldo Moro 5, 00185 Rome, Italy}
\address[11]{INAF -- Osservatorio Astronomico d'Abruzzo, Via M. Maggini snc, I-64100, Teramo, Italy}
\address[12]{School of Physical Science and Technology, Southwest Jiaotong University, Chengdu, 610031, China}

\abstract{  
This work shows how human physical reasoning can guide machine-driven symbolic regression toward discovering empirical laws from observations. As an example, we derive a simple equation that classifies fast radio bursts (FRBs) into two distinct Gaussian distributions, indicating the existence of two physical classes. This human–AI workflow integrates feature selection, dimensional analysis, and symbolic regression: deep learning first analyzes CHIME Catalog 1 and identifies six independent parameters that collectively provide a complete description of FRBs; guided by Buckingham–$\pi$ analysis and correlation analysis, humans then construct dimensionless groups; finally, symbolic regression performed by the machine discovers the governing equation. When applied to the newer CHIME Catalog, the equation produces consistent results, demonstrating that it captures the underlying physics. This framework is applicable to a broad range of scientific domains.}



\makeatletter
\renewcommand{\@authoremail}{
Yu Wang, email: yu.wang@icranet.org;
Wenbin Lin, email: lwb@usc.edu.cn;
Bing Zhang, email: bzhang1@hku.hk
}
\makeatother
\maketitle
    
\begin{multicols}{2}

\section{Introduction}\label{sec:intro}

Phenomenology constructs approximated equations that predict and explain empirical data without requiring exhaustive theoretical foundations, bridging experimental observations and theoretical laws \cite{husserl2001phenomenology, moran2002introduction}. Machine learning (ML) \cite{2015Sci...349..255J} is well-suited for phenomenological studies because it can identify patterns and correlations from data \cite{2019RvMP...91d5002C}.

Symbolic regression \cite{makke2024interpretable}, a branch of machine learning, searches for mathematical expressions that best fit the data \cite{fortin2012deap, 2023arXiv230501582C,2024arXiv240419756L}. For example, it rediscovered Kepler’s third law \cite{2023NatCo..14.1777C} and Newton’s laws \cite{2023MLS&T...4d5002L}. It modeled the Kondo temperature in condensed matter systems \cite{2021PhRvB.104w5111M} and recovered the Lorenz system equations of dynamical systems \cite{2024CSF...18815538D}. For further examples, we refer to review papers such as \cite{2022NatRP...4..399K, angelis2023artificial, makke2024interpretable}.

Despite these successes, symbolic regression still faces fundamental limitations. Most studies validate their methods by rediscovering known laws with a few parameters rather than discovering new, especially complex, ones. This is partly because the search space grows combinatorially with the size of the operator set, the expression depth, and the number of parameters. The presence of noisy data further increases computational cost and produces many near-equivalent expressions.

Therefore, the machine needs human experience and guidance. We establish a machine phenomenology framework in which human reasoning and machine learning are coupled to derive empirical equations from observational data. The workflow proceeds as follows. First, deep learning is used for feature selection, identifying a minimal set of independent parameters that capture the essential variability of the system. Second, guided by dimensional analysis and the Buckingham–$\pi$ theorem, humans construct dimensionless groups that encode physical invariances and scaling relations. Third, these dimensionless quantities are provided to the machine, which performs symbolic regression to search for analytical expressions that best describe the data. Finally, the resulting equations are cross-validated on independent datasets to test their robustness and generality.

We apply this procedure to the classification of fast radio bursts (FRBs) \cite{lorimer2007bright,thornton2013population,2023RvMP...95c5005Z}. 

FRBs are intense, millisecond-duration radio flares that predominantly originate at cosmological distances, making them one of the most compelling unsolved mysteries in contemporary astrophysics \cite{2018Katz_FRB, Popov_2018, 2019Cordes_FRB, Petroff2019,2023RvMP...95c5005Z}. Since their first detection in 2007 by the Parkes Telescope \cite{2007Lorimer}, thousands of FRBs have been identified, but their origins remain elusive. These enigmatic bursts are observationally divided into repeaters (multiple bursts) and non-repeaters (single observed bursts), though some non-repeaters may simply be under-observed. Whether these two types are intrinsically distinct in their physical origins remains an active and debated question \cite{ai2021true}.  Before testing the physical models that explain the characteristics of these two potential FRB types, it is essential first to confirm the existence of this dichotomy. We aim to find an equation from the observables, which is accurate and interpretable, to verify the possible FRB classes.

This article is structured as follows. The selection and analysis of FRB parameters are discussed in Section \ref{sec:param}. Two different symbolic regression approaches to deriving classification equations, along with their results, are presented in Section \ref{sec:equations}. Section \ref{sec:results} reports our results and the application of the discovered equations to new data. The final discussion and conclusions can be found in Section \ref{sec:conclusion}.

\section{Parameter Selection and Analysis}
\label{sec:param}

\subsection{CHIME Catalogs}
\label{sec:data}


We adopted data from the first CHIME/FRB catalog \cite{chime/frbcollaboration2021FirstCHIMEFRB}, which included 536 FRBs observed between July 25, 2018 and July 1, 2019. This dataset contains 474 bursts from non-repeating sources and 62 bursts from 18 repeating sources. Hereafter, we refer to the first CHIME/FRB catalog as Catalog 1.

We treated each sub-burst as an independent event, resulting in a total of 600 sub-bursts \cite{2022zhuge_nosup}. To ensure data consistency, we excluded six non-repeating bursts\footnote{FRB20190307A, FRB20190307B, FRB20190329B, FRB20190329C, FRB20190531A, FRB20190531B} that lack flux. This pre-processing step produced a final dataset of 594 independent sub-bursts, with 500 from apparent non-repeaters and 94 from repeaters.

We note that due to instrumental bandwidth limitations, the values of certain parameters are constrained, and the corresponding entries in the catalog may not be entirely accurate. Nevertheless, we retain these data points during model training. This decision is motivated by, first, given the limited sample size, excluding these data would result in the removal of approximately half of the non-repeating sources. Second, such points can be interpreted as statistical upper or lower limits, which offer informative constraints that the model may still exploit.

The second catalog \cite{chime20repeater}, identified between September 30, 2019 and May 1, 2021, consists of 98 high confidence bursts, totaling 119 sub-bursts from 25 repeating sources.  Six of these had previously been classified as non-repeaters in Catalog 1 \cite{chime/frbcollaboration2021FirstCHIMEFRB}. We will use this new catalog to evaluate our derived equations. The second catalog also includes 14 uncertain candidates, which are not included in our analysis.  Hereafter, we call the second catalog as Catalog 2. This catalog is an updated repeater catalog, and the complete second catalog is still underway.


When required by symbolic regression, for example, in the parameter selection procedure, we normalize the data set using Z-score normalization\cite{2015arXiv150306462G}. This is achieved by transforming the data using the formula  
\begin{equation}
X^{'} = \frac{X - \mu}{\sigma}
\end{equation}  
where $\mu$ is the mean and $\sigma$ is the standard deviation of the data $X$. Normalization of the Z-score ensures that the transformed data retain the shape of the original distribution while aligning its mean and standard deviation to 0 and 1, respectively. This normalization helps improve numerical stability and accelerates convergence by preventing features with larger magnitudes from dominating the learning loss. 



Concerning the imbalance of data samples between non-repeaters and repeaters, with the repeaters being roughly five times less frequent. It is essential to prevent our models from inadvertently learning this imbalance. To address this issue, we utilized the Synthetic Minority Over-sampling Technique (SMOTE) \cite{chawla2002SMOTESyntheticMinority}, as implemented in the \texttt{imbalanced-learn} package \cite{lemaitre2017ImbalancedlearnPythonToolbox}, to artificially expand the repeater sample, matching its size to the non-repeater sample in the training set. As a result, after augmentation, the training set includes an equal number of repeaters and non-repeaters.

The formulas derived from symbolic regression are orders of magnitude simpler compared to normal neural networks, and we tend to prefer simpler formulas. Therefore, overfitting in symbolic regression is much less severe than the normal neural networks. In fact, these formulas might even overlook some details. Hence, when performing symbolic regression, it is not necessary to strictly separate the training and testing datasets. In our case, we use all data from Catalog 1 to derive the formulae, which are subsequently applied to both Catalog 1 and 2.

\subsection{Six Key Parameters}
\begin{figure}[H]
    \centering
    \includegraphics[width=1\linewidth]{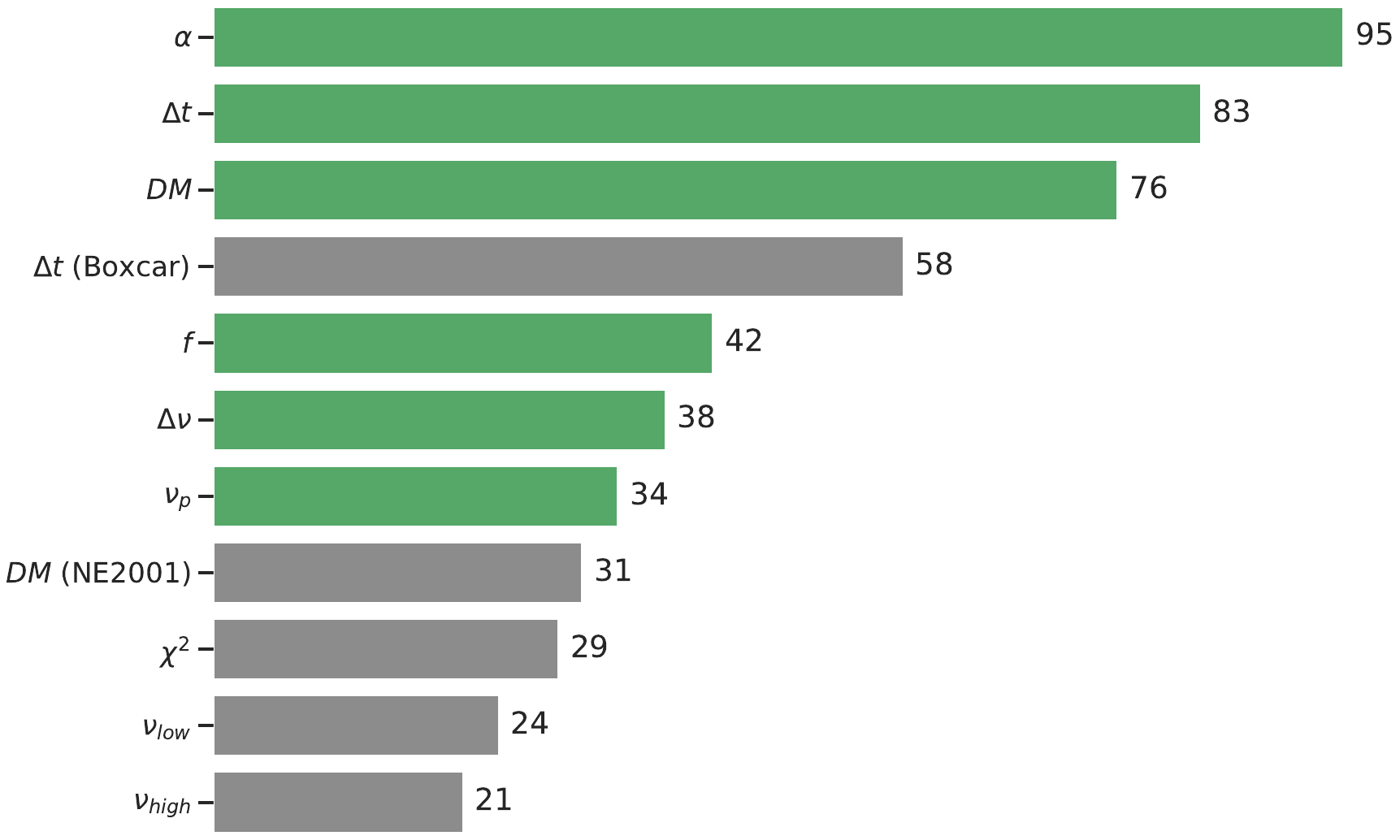}
    \caption{Visualization of the feature selection. The plot shows the frequency with which each feature was ranked among the top six most important features across 100 neural network models. The green rectangles highlight the six selected features. Parameters include: $\alpha$: spectral index; $\Delta t$: sub-burst width; $DM$: excess DM using YMW16 model; $\Delta t$ (Boxcar): entire burst width from boxcar method; $f$: flux density; $\Delta \nu$: frequency bandwidth; $\nu_p$: peak frequency; $DM$ (NE2001): excess DM using NE2001 model; $\chi^2$: chi-square statistic; $\nu_{low}$: lower frequency bound; $\nu_{high}$: upper frequency bound.}
    \label{fig:select_feature2}
\end{figure}

The Catalog 1 lists 48 observational parameters\footnote{For a detailed description of all the parameters, we refer to \cite{chime/frbcollaboration2021FirstCHIMEFRB} and \cite{chime20repeater}.}. To identify the most critical parameters, we first perform parameter selection (feature selection). Specifically, we train 100 fully connected neural networks, each with four hidden layers and a total of 33,280 parameters, using  all 48 observational parameters to classify non-repeaters and repeaters. To assess the importance of each observational parameter, we use SHAP (Shapley Additive Explanations) values \cite{NIPS2017_7062}, which quantify how much each feature contributes to the classification results. For each network, we rank the six most important parameters based on their SHAP values. Figure \ref{fig:select_feature2} shows the total count of each parameter, representing how often it appears among the top six across all networks. Finally, we select the most frequently occurring six parameters. Notably, we include $\Delta t$ but not $\Delta t$ (Boxcar). This choice is based on our analysis method, which examines each sub-burst individually. The $\Delta t$ parameter measures the duration of each single sub-burst, making it suitable for our method, while $\Delta t$ (Boxcar) represents the total duration of an entire burst that may contain multiple sub-bursts. The boxcar width typically includes additional intra-channel dispersion smearing and scattering, which do not robustly represent the intrinsic pulse width \cite{chime/frbcollaboration2021FirstCHIMEFRB,sun2025exploring}.

\begin{figure*}[htbp!]
    \centering
    \includegraphics[width=0.8\linewidth]{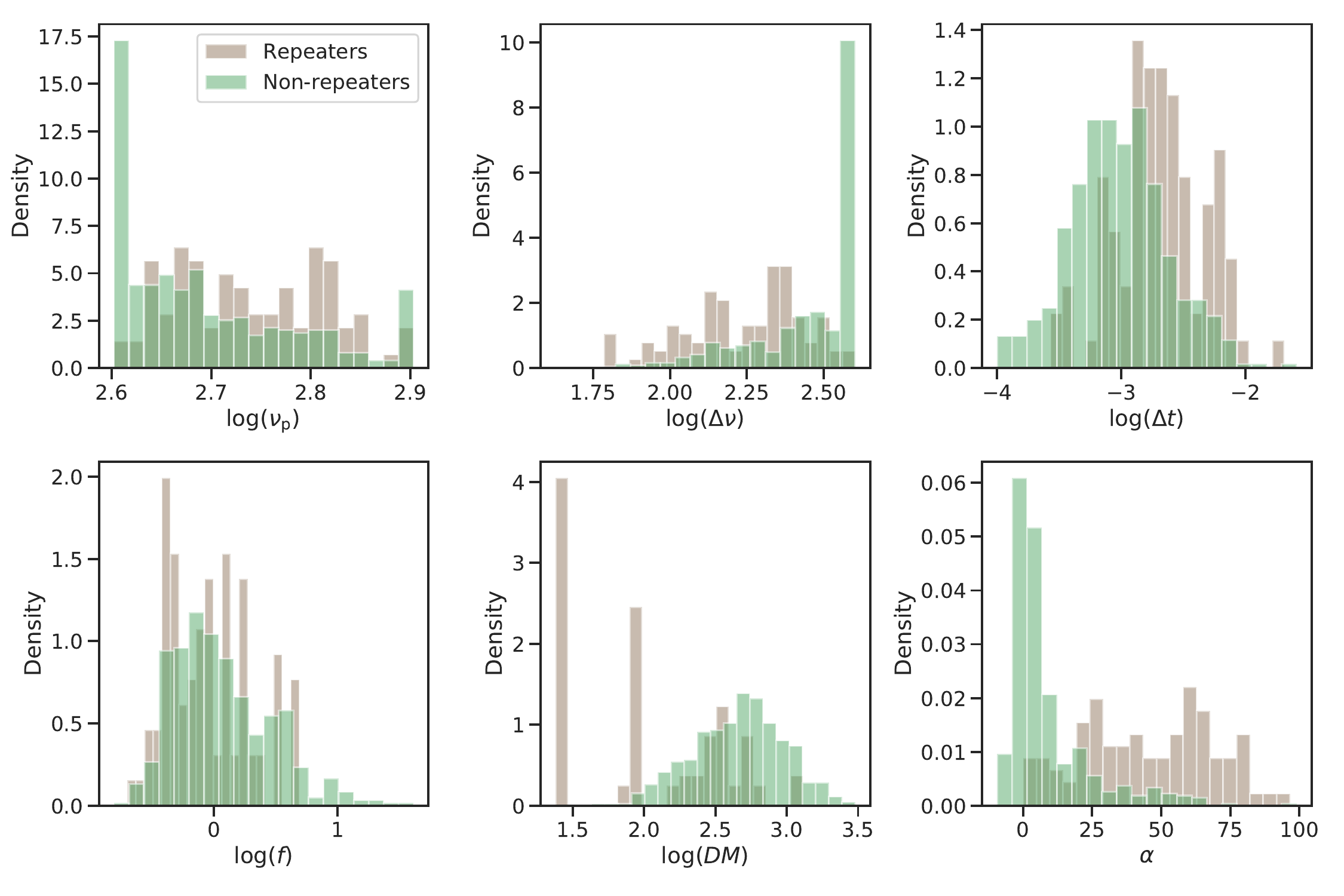}
    \caption{Input features distributions for repeating and non-repeating FRBs. The peaks at the edges of the $\nu_p$ and $\Delta_\nu$ distributions originate from the limited bandwidth of CHIME, while the peak in the $DM$ distribution arises from multiple bursts produced by repeaters.}
    \label{fig:para_dis}
\end{figure*}

\begin{figure*}[htbp!]
    \centering
    \includegraphics[width=0.9\linewidth]{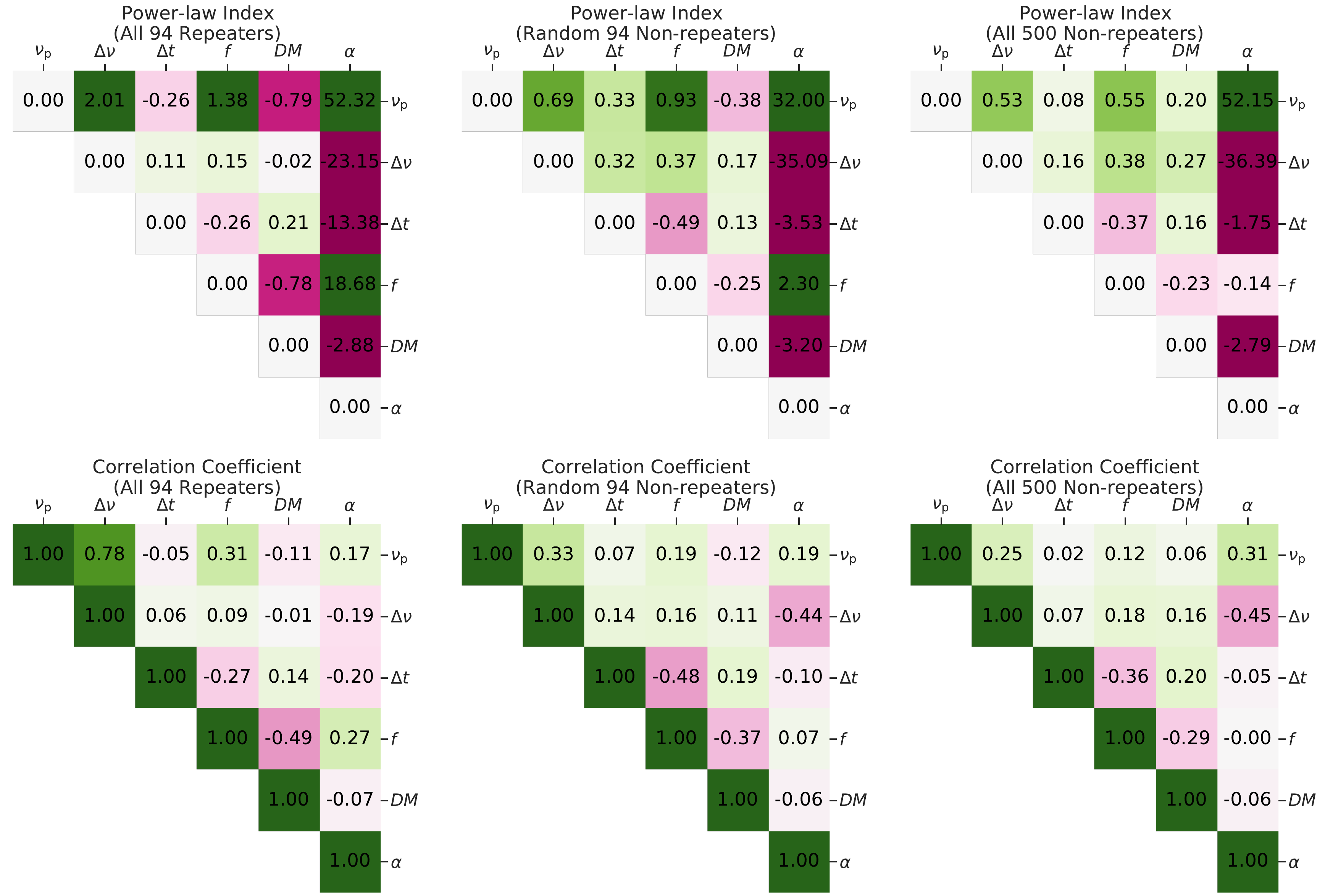}
    \caption{Parameter relationships between repeating and non-repeating FRBs, with numerical values annotated in each cell. The top row shows power-law indices and the bottom row shows correlation coefficients between six  parameters. Left panels show results for all repeating FRBs, middle panels show a random sample of non-repeaters matched to the repeater sample size, and right panels show results for all non-repeating FRBs.}
    \label{fig:relation}
\end{figure*}


These six parameters selected by the neural network exactly provide a comprehensive description of FRB observations, with each parameter playing a unique role that cannot be substituted by another. From an observational perspective, $\Delta \nu$ (\texttt{freq\_width = high\_freq - low\_freq})\footnote{The variable names used in the CHIME/FRB catalog are shown in parentheses.} represents the energy range of the spectrum. $\nu_p$ (\texttt{peak\_freq})  and $\alpha$ (\texttt{sp\_idx}) specify the peak frequency and  the spectral index\footnote{In CHIME/FRB catalog, the spectral energy distribution is empirically described by
$f(\nu) \propto \left(\frac{\nu}{400\,\text{MHz}}\right)^{\gamma + r\ln\left(\frac{\nu}{400\,\text{MHz}}\right)}$,
where $\gamma$ (\texttt{sp\_idx}) is the spectral index and $r$ (\texttt{sp\_run}) is the spectral running. The spectral index $\gamma$ thus defines the local logarithmic slope of the spectrum at the frequency of 400 MHz \cite{2021ApJ...923....1P}.}, respectively, together they define the spectral shape. $f$ (\texttt{flux}) measures the flux density, representing the intensity of the radiation, and $\Delta t$ (\texttt{width\_fitb}) describes the duration of the emission. Mathematically, these parameters define a spectrum over a specific energy range and duration. Additionally, $DM$ (\texttt{dm\_exc\_ymw16}) estimates the source’s distance, enabling the conversion of observed quantities into the cosmological rest frame for inferring the physical properties of the FRB. The distributions of the six parameters are shown in Figure \ref{fig:para_dis}.

We test the neural network using these six parameters, it yields an accuracy of $\sim 95\%$. This accuracy sets a benchmark for our symbolic regression approach. In theory, a sufficiently complex function could achieve comparable accuracy. However, in practice, we prioritize simplicity alongside accuracy, which means the final equation may fall slightly short of this benchmark. This trade-off is deliberate, as our equations are designed to normally include fewer than 10 symbolic operations by these selected parameters, in contrast to the neural network’s 33280 parameters (weights of neural network) performing numerous nonlinear computations.

\subsection{Physical Implication From Selection}
First, these six parameters collectively represent a complete set that captures the diverse observational characteristics of FRBs. Their ability to describe spectral shape, burst duration, brightness, and distance-related features indicates that the differences between repeating and non-repeating bursts are not confined to a single aspect but extend across the full range of observable properties. This finding supports the hypothesis that the two classes may arise from distinct processes.

Second, these six parameters are selected by the machine from a larger set of 48 observational parameters. Their coherent emergence statistically suggests that the differences between repeaters and non-repeaters are intrinsic rather than a result of observational bias.

\subsection{Parameters Pairwise Relations} \label{sec:param-relation}

We compute the pairwise relationships among six parameters and fitted the data points of repeaters and non-repeaters separately using a power-law model, obtaining the power-law index. We also calculate the correlation coefficient for each relation \footnote{Since the spectral index $\alpha$ is itself an exponent, its power-law fitting is performed using $\exp(\alpha)$}. The results are shown in Figure \ref{fig:relation}. Our analysis included all 94 repeaters and 500 non-repeaters. Furthermore, to account for data randomness and to better compare with the statistics of repeaters, we randomly sampled 94 non-repeaters.

\begin{figure}[H]
    \centering
    \includegraphics[width=1\linewidth]{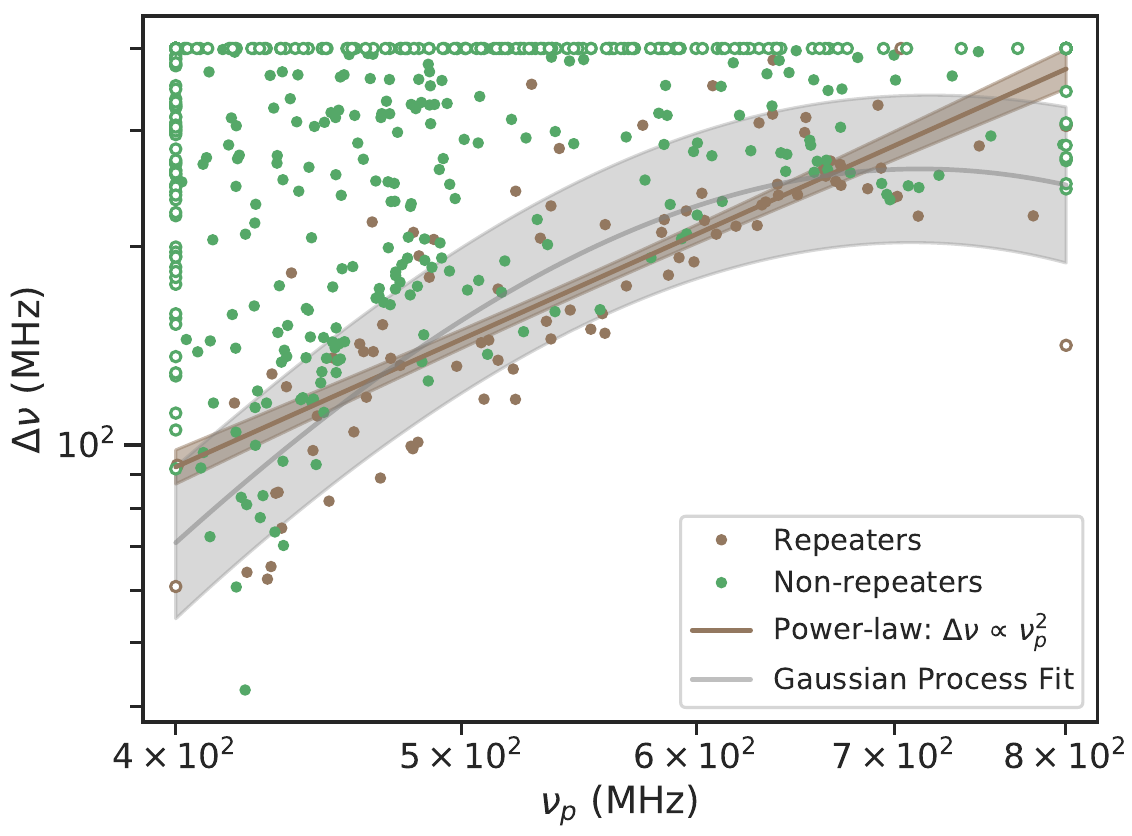}
    \caption{Relationship between peak frequency ($\nu_p$) and frequency width ($\Delta_\nu$) for repeating (brown dots) and non-repeating (green dots) FRBs. A power-law fit for repeaters (brown line) shows the scaling relationship $\Delta_\nu \propto \nu_p^2$, with the shaded region representing the 1-$s\sigma$ uncertainty region. A non-parametric Gaussian Process fit (grey line) with its 1-$\sigma$ uncertainty region is also shown, capturing potential nonlinear trends in the relationship. The data points hit the boundaries, marked by dots with white centers, are due to the limited bandwidth of the CHIME telescope.}
    \label{fig:vp-dv-relation}
\end{figure}

The strongest correlation for repeaters is $\nu_p$ vs $\Delta\nu$ (0.78), with a steep positive slope (2.01), plotted in Figure \ref{fig:vp-dv-relation}. For non-repeaters (both full and random subsets), this correlation is weaker (0.25–0.33) with flatter slopes (0.53–0.69). The strong positive correlation suggests that bursts with higher peak frequencies tend to have wider frequency ranges. This could indicate that the emission mechanism produces broader spectra at higher frequencies, possibly due to the energy distribution of emitting particles or/and the size of the emission region. In addition, many repeaters exhibit down-drifting \cite{hessels2019frb,zhou2022fast}, where later emissions occur at progressively lower frequencies. This suggests a systematic relation between $\nu_p$ and $\Delta \nu$, strengthening this correlation. 

The anti-correlation between  $f$  and  $DM$  is stronger in repeaters ($-0.49$) than in non-repeaters ($-0.29$), supporting the idea that flux diminishes with distance (proportional to  $DM$ ) due to the inverse-square law. The weaker trend in non-repeaters may reflect a combination of progenitor distance variations and the broad intrinsic luminosity distribution.

Non-repeaters exhibit a moderate anti-correlation ($-0.45$) between  $\alpha$  and  $\Delta \nu$, indicating that broader bandwidths are associated with flatter spectra. In contrast, repeaters show no significant trend ($-0.190$).

Both populations exhibit negative Pearson correlations between  $\Delta t$  and  $f$  (repeaters: $-0.27$; non-repeaters: $-0.36$), implying that brighter bursts tend to release energy over shorter timescales. The weaker correlation in repeaters may suggest that their energy release is more stochastic.

However, we must also note that the above discussions may be related to the sample size. Although there are a total of 94 bursts from repeaters, they originate from only 18 sources. Statistically, this may lead to conclusions suggesting similar origins. Further observations are needed to assess the significance of the observational effect.

\begin{figure*}[htbp!]
    \centering
    \includegraphics[width=0.8\linewidth]{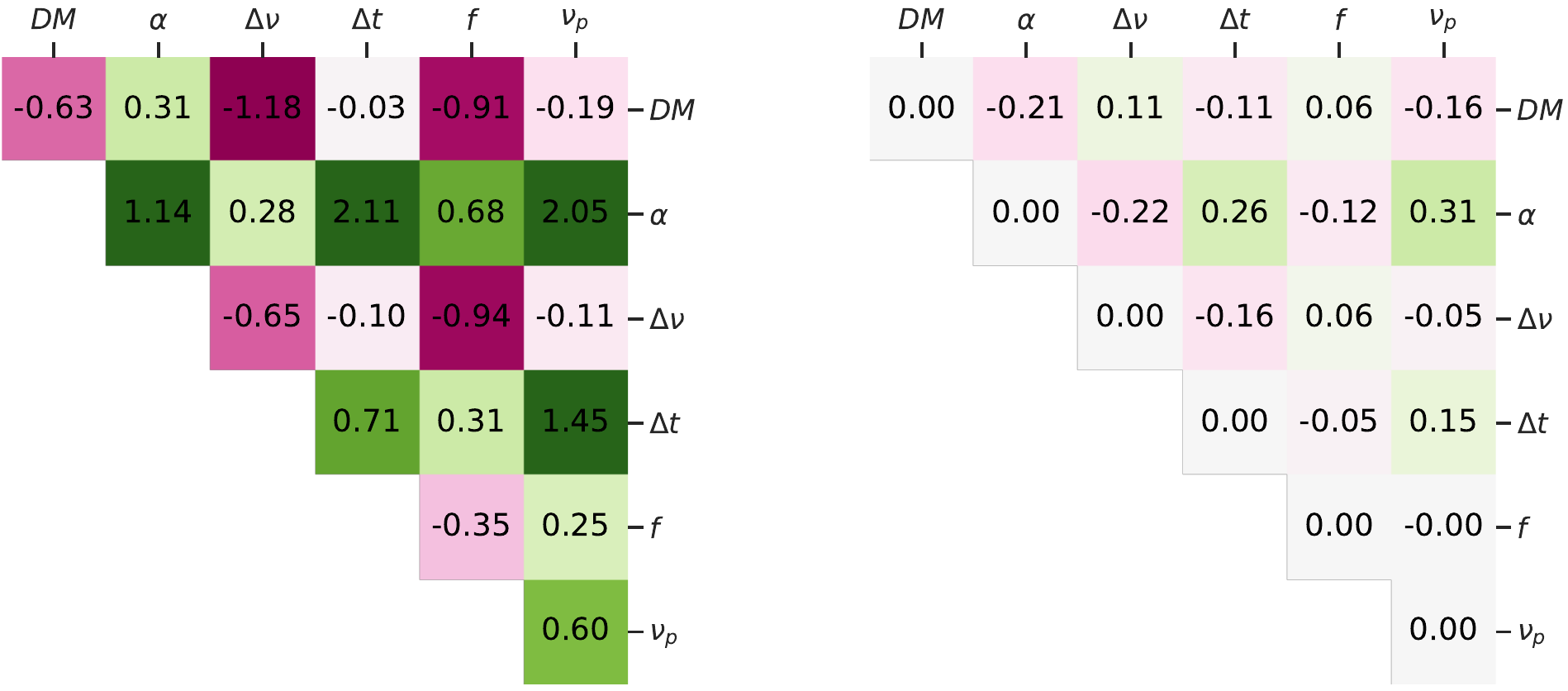}
    \caption{\textbf{Left}: Interaction matrix of $\Delta g_{x_i}$ (diagonal elements) and $\Delta g_{x_i, x_j}$ (off-diagonal elements), indicating the importance of the parameters. \textbf{Right}: Interaction matrix of $\eta_{x_i, x_j}$, indicating the nonlinear relationships.}
    \label{fig:interactions}
\end{figure*}

\subsection{Parameters Importance and nonlinearity} \label{sec:analyzing}

Understanding the nonlinearity of input features (six parameters) and their interactions is helpful for guiding equation discovery in symbolic regression. To address this, we develop a perturbation framework that quantifies feature importance and pairwise interactions by analyzing output deviations under input variations. 

The sensitivity of a model $ g $ to an individual feature $ x_i $ is determined by first computing the baseline output
\begin{equation}
g_0 = g(x_1, x_2, \ldots, x_i, \ldots, x_m).
\end{equation}
Each feature $ x_i $ is then perturbed by an increment $ \delta x_i = \sigma_{x_i} $, where $ \sigma_{x_i} $ represents the standard deviation of $ x_i $ across the dataset. The perturbed output
\begin{equation}
g_{x_i} = g(x_1, x_2, \ldots, x_i + \delta x_i, \ldots, x_m)
\end{equation}
is then used to calculate the feature’s influence through the first-order deviation 
\begin{equation}
\Delta g_{x_i} = g_{x_i} - g_0.
\end{equation}
Features exhibiting larger absolute $ |\Delta g_{x_i}| $ values influence the model’s prediction the most, as their perturbation induces the largest output shifts.  

To disentangle pairwise interactions between features $ x_i $ and $ x_j $, both parameters are perturbed simultaneously ($ x_i + \sigma_{x_i}, x_j + \sigma_{x_j} $), resulting in the joint perturbed output
\begin{equation}
g_{x_i, x_j} = g(x_1, x_2, \ldots, x_i + \delta x_i, \ldots, x_j + \delta x_j, \ldots, x_m).
\end{equation}
The combined effect is quantified as
\begin{equation}
\Delta g_{x_i, x_j} = g_{x_i, x_j} - g_0,
\end{equation}
representing the importance of changing a pair of parameters simultaneously.

To identify nonlinear interactions between pairs of parameters, we introduce
\begin{equation}
\eta_{x_i, x_j} = \Delta g_{x_i, x_j} - \Delta g_{x_i} - \Delta g_{x_j}
\end{equation}
to isolate interaction effects that deviate from additive contributions. A non-zero $ \eta_{x_i, x_j} $ signals the presence of nonlinear dependencies.

This methodology aids in hypothesizing the structure of potential equations for symbolic regression. By analyzing $\Delta g_{x_i}$ and $\Delta g_{x_j}$ along with $\eta_{x_i, x_j}$, we can infer possible relationships. Some examples are: 1) When opposing signs in $ \Delta g_{x_i} $ and $ \Delta g_{x_j} $ coincide with $ \eta_{x_i, x_j} \approx 0 $, the interaction is consistent with linear subtraction. 2) A positive $ \eta_{x_i, x_j} $, e.g., $ \Delta g_{x_i} = +0.3 $, $ \Delta g_{x_j} = +0.2 $, $ \eta_{x_i, x_j} = +0.3 $, indicates amplification beyond additive effects. This aligns with multiplicative terms like $ x_i \cdot x_j $, where perturbing both features introduces cross-terms. 3) Non-zero but small $ \eta_{x_i, x_j} $ suggests higher-order interactions, e.g. $ g \propto (x_i + x_j)^2 $ introduces a cross-term $ 2x_i x_j $, which amplifies $ \eta $ when both features are perturbed. 

Figure \ref{fig:interactions} shows the interaction matrices of the six parameters, which will be implemented in the next section.

\section{Methods}
\label{sec:equations}

Two approaches are employed in our phenomenological formula search process. The first approach is to fit an equation that is a product of multiple power laws, which is a common form in astrophysics. The second approach involves non-dimensionalization by human first, followed by the machine identifying a classification equation.

\subsection{Criteria: Precision, Recall and F2 Score}

In FRB classification, sources are divided into repeaters and non-repeaters. However, due to limited observational coverage and inherent uncertainties, some sources currently classified as non-repeaters may eventually be recognized as repeaters. Moreover, repeaters are substantially fewer in number compared with non-repeaters for current samples \cite{chime/frbcollaboration2021FirstCHIMEFRB, chime20repeater}.

To address these issues, we adopt the F2 score as our evaluation metric \cite{sasaki2007truth}. The F2 score is defined as a weighted harmonic mean of precision and recall. Precision is given by
\begin{equation}
P = \frac{\text{TP}}{\text{TP} + \text{FP}},
\end{equation}
where TP denotes the true repeaters correctly identified and FP denotes the non-repeaters misclassified as repeaters. Recall is defined as
\begin{equation}
R = \frac{\text{TP}}{\text{TP} + \text{FN}},
\end{equation}
with FN representing repeaters that are missed. The generalized F-score is formulated as
\begin{equation}
F_\beta = (1 + \beta^2) \cdot \frac{P \times R}{\beta^2 \cdot P + R}.
\end{equation}
Setting $\beta = 2$ weights recall more heavily than precision, which is important when missing a true repeater is considered a more serious error than including some non-repeaters. In the meanwhile, it minimizes the false alarm that the non-repeaters incorrectly labeled as repeaters.

Our equation searching process prioritizes optimizing recall first, followed by the F2 score. Given the class imbalance in the dataset (500 $vs$ 94) and considering each class should be equally important, we first compute the statistical evaluation metrics respectively for each class and then take the average.

\subsection{Power-Law Multiplication}
\label{sec:pls}
In astrophysics, equations frequently take the form of power-laws, which can be understood from both theoretical and observational perspectives. From the theoretical aspect, power-laws naturally emerge in systems that exhibit scale-invariant behavior, as the same physics operates across different scales spanning orders of magnitude in size, time and energy. Observations in astrophysics often span limited ranges of data, and power-laws are simple models that fit well to log-log plots, where a straight line indicates a power-law relationship. 

Coherent emission processes, thought to be the FRB emission mechanism, often produce power-law spectra. Dispersion and scattering in the interstellar medium, which affect FRBs, often follow power-laws due to the frequency dependence of plasma interactions. Such physical mechanisms are common in FRB studies, and the observations, such as the energy distribution, fluence, and event rates are often analyzed using power-law fits.

Therefore, the frequent appearance of power-laws in astrophysical equations, especially in FRBs, reflects both the fundamental nature of many astrophysical processes and the practical utility of these models in interpreting observational data.  Hence, it is reasonable to assume that the threshold differentiating repeaters and non-repeaters can be approximately described by the multiplication of power-laws,
\begin{equation}
    \Theta \propto \prod_{i} X_i^{-\beta_i},
\end{equation}

We first applied transformations to the 6 selected parameters. For parameters expected to follow power-law scaling ($f, DM, \Delta \nu, \Delta t, \nu_p$), we used a logarithmic transformation, $x_i = \log_{10}(X_i)$. However, the spectral index $\alpha$ is already an exponent and can take negative values, we included $\alpha$ directly. This converts the multiplicative relationships into the additive form which enabled us to apply the linear regression models
\begin{equation}
\theta = \beta_0 + \beta_1 x_1 + \beta_2 x_2 + \dots + \beta_6 x_6 
\label{eq:linef}
\end{equation}
where $\theta = \log_{10}(\Theta_{\rm PL})$ is the dependent variable (e.g., potential physical quantity to determine if the FRB belongs to a repeat burst), $ x_1, x_2, \dots, x_6 $ are the selected key parameters, $ \beta_0 $ is a constant, $ \beta_1, \beta_2, \dots, \beta_6 $ are the regression coefficients that capture the influence of each parameter. This approach helps to construct a simplified model that captures FRB phenomena and identifies patterns that are not immediately apparent in the raw data.

\subsection{Neural Dimensionless Regression (NDR)}
\label{sec:ndr}
\begin{figure*}[htbp!]
    \centering
    \includegraphics[width=0.8\linewidth]{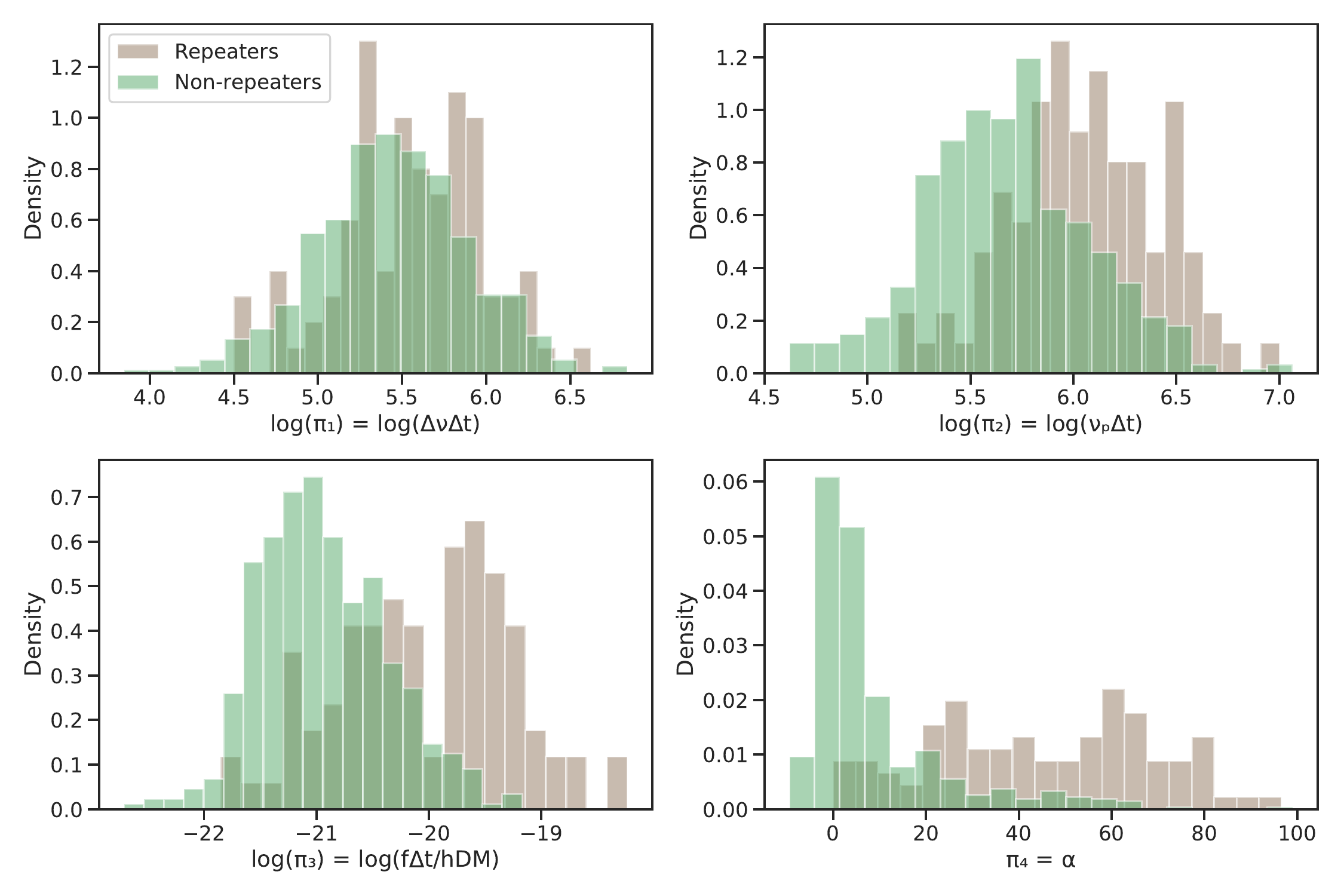}
    \caption{Distribution of dimensionless groups comparing repeating (blue) and non-repeating (green) FRBs.}
    \label{fig:Distribution_dimensionless}
\end{figure*}

Normally the first step of neural network training is normalization, features are normalized because datasets with large variations may lead to numerical instability and network divergence. However, normalization cancels out the units of physical quantities, causing the network to lose information about their dimensional relationships. Consequently, when deriving equations, the machine may produce physically incorrect formulas.

Currently, symbolic regression methods address dimensional consistency by incorporating additional constraints during training. For instance, PySR does not inherently enforce unit consistency, so users must manually define constraints, such as restricting operations to guide the regression process or filtering out results that do not match expected units after training \cite{2023arXiv230501582C}. In contrast, PhySO requires users to specify the units of each input and output variable, it enforces unit consistency by allowing only equations where all terms have matching units \cite{PhySO_RL_DA}. Ensuring dimensionally correct equations in these methods comes at a high computational cost, particularly for complex systems. And the enforcement of dimensional consistency may exclude potential candidate equations, especially when minor deviations arise due to limitations in data precision, restricting their applicability to noisy data which is common in astronomy. 

We propose a human-machine collaborative approach, where humans perform physics-informed preprocessing to produce dimensionless features and reduce the number of features using dimensional constraints. Consequently, the machine no longer needs to account for dimensional consistency but already takes advantage of dimensional analysis from the human’s preprocessing, thereby reducing computational time and mitigating the risk of missing the correct equation due to the enforcement of dimensional consistency in the presence of noisy data. We refer to this method as the \textit{Neural Dimensionless Regression} (NDR).

The mathematical foundation of NDR is the Buckingham $\pi$ theorem \cite{bertrand1878homogeneite}, which states that any physical equation involving  $n$  variables and  $k$  fundamental physical dimensions can be rewritten in terms of  $(n - k)$  independent dimensionless groups called $\pi$ groups.

The first step in applying the $\pi$ theorem is to list the relevant variables and their corresponding physical dimensions. We consider six measured parameters along with one physical constant: Planck’s constant ($h$). The fundamental dimensions involved are mass [$M$], length [$L$], and time [$T$], giving a total of six dimensional variables and one dimensionless variable ($\alpha$). The physical dimensions of these features are summarized as follows

\begin{enumerate}
    \item Spectral width ($\Delta \nu$): $[T^{-1}]$
    \item Peak frequency ($\nu_p$): $[T^{-1}]$
    \item Flux density ($f$): $[M L^{-2}]$
    \item Duration ($\Delta t$): $[T]$
    \item Dispersion measure (DM): $[L^{-2}]$
    \item Spectral index ($\alpha$): Dimensionless
    \item Planck’s constant ($h$): $[M L^2 T^{-1}]$
\end{enumerate}

Since there are three fundamental dimensions [$M, L, T$], the number of independent dimensionless groups is determined by the Buckingham $\pi$ theorem as $n - k = 6 - 3 = 3$, meaning three independent dimensionless groups, plus the already dimensionless $\alpha$.

To construct these groups, it is necessary to select three repeating variables that must satisfy two conditions
\begin{itemize}
    \item They must span all fundamental dimensions [$M, L, T$];
    \item They must be physically independent, meaning that no variable can be derived from the others. 
\end{itemize}
However, in practice, there is more than one set of variables that meet these two conditions, which is one of the challenges of dimensional analysis. In section \ref{sec:analyzing}, we have introduced the perturbation method, which can assist in selecting the optimal set of repeating variables in at least two aspects
\begin{itemize}
    \item Importance: variables with high influence (high $\Delta g$) are likely to have a larger impact on the system and may be better candidates as repeating variables;
    \item Simplicity: variables with low linearity (low $\eta$) indicate less synergistic or antagonistic interaction, ensuring variables contribute distinct physical effects to the system and simplifying the dimensionless groups to produce scaling laws that are more intuitive and generalizable.

\end{itemize}

Among the six original parameters, only the flux ($ f $) contains the mass dimension $[M]$. To make it dimensionless, we have introduced a seventh parameter, the Planck constant $ h $, which also carries the $[M]$ dimension. Since we are studying electromagnetic radiation, the introduction of the Planck constant is also physically meaningful. Both $ h $ and $ f $ encompass the three fundamental dimensions [$M, L, T$] and are mathematically eligible to be selected together in a dimensionless group. However, perturbation method results (Figure \ref{fig:interactions}, left) show that among the six parameters, $ f $ has the least influence on the outcome. Therefore, to introduce the $[M]$ dimension, we select either $ h $ or $ f $, while choosing parameters with greater influence for the other dimensions. Since selecting $ h $ or $ f $ does not affect the dimensionless group, as both will appear in the same group, we choose $ h $ here.  

Among the remaining five parameters, $ \alpha $ is inherently dimensionless, and $ DM $ is associated with the $[L]$ dimension, while the other three parameters correspond to the $[T]$ dimension. Since $ DM $ is the only parameter related to $[L]$, it is a necessary choice. Additionally, perturbation method results indicate that $ DM $ itself has a high level of importance.  

For selecting the parameter representing the $[T]$ dimension, we first evaluate the importance of the three parameters: $ \Delta_{\Delta t} > \Delta_{\Delta \nu} > \Delta_{\nu_{p}} $. Next, we analyze the degree of nonlinear interactions between these parameters and the others (Figure \ref{fig:interactions}, right). 
Comparing the $ \eta $ values of these three variables with respect to $ DM $ and $ f $, which represent the other two dimensions, we find $ \eta_{\Delta t} = \eta_{\nu_{p}} < \eta_{\Delta \nu} $. Consequently, we select $ \Delta t $ as the representative parameter for the $[T]$ dimension.

Finally, the choice is $\{ h, \text{DM}, \Delta t \}$,  these variables collectively provide independent units of mass (via $h$), length (via DM and $h$), and time (via $\Delta t$ and $h$). Each remaining variable is then expressed as a dimensionless product involving powers of these repeating variables.

\begin{table}[h]
    \centering
    \renewcommand{\arraystretch}{1.4}  
    \begin{tabular}{@{} l c c @{}}
        \toprule
        \textbf{Term} & \textbf{Expression} & \textbf{Log Form} \\
        \midrule
        $\pi_1$ & $ \Delta \nu \Delta t $ & $\log(\texttt{width\_fitb}) + \log(\texttt{fre\_width})$ \\
        $\pi_2$ & $ \nu_p \Delta t $ & $\log(\texttt{peak\_freq}) + \log(\texttt{width\_fitb})$ \\
        \multirow{2}{*}{$\pi_3$} & \multirow{2}{*}{$ \frac{f \Delta t}{h ~ \text{DM}} $} & $ \log(\texttt{flux}) + \log(\texttt{width\_fitb}) $ \\
         & & $ -\log(\texttt{dm\_exc\_ymw16 }) - \log(h)$ \\
        $\pi_4$ & $ \alpha $ & $\texttt{sp\_idx}$ \\
        \bottomrule
    \end{tabular}
    \caption{Dimensionless groups and their logarithmic representations. The logarithmic form is expressed using the Catalog's variables. The spectral index $\alpha$ remains in its original form since it is already an index.}
    \label{tab:pi_groups}
\end{table}

The first dimensionless parameter is obtained by combining $\Delta \nu$ and $\Delta t$:
\begin{equation}
\pi_1 = \Delta \nu  \Delta t,
\end{equation}
this combination is the time-bandwidth product. A larger value might indicate more dispersed signals or certain emission mechanisms. 

The second dimensionless parameter is constructed using $\nu_p$ and $\Delta t$:
\begin{equation}
\pi_2 = \nu_p  \Delta t,
\end{equation}
this term quantifies temporal-spectral correlation, could be relate to the number of wave cycles in the burst. 

Constructing a dimensionless parameter involving flux $f$:
\begin{equation}
\pi_3 = \frac{f  \Delta t}{h  \text{DM}},
\end{equation}
this term compares observed flux to intrinsic energy density, scaled by plasma dispersion effects. A higher values suggest more luminous bursts.

The spectral index $\alpha$ is already dimensionless and remains as 
\begin{equation}
\pi_4 = \alpha,
\end{equation}
which characterizes spectral shape.


The application of the Buckingham $\pi$ theorem thus reduces six dimensional parameters to four independent dimensionless terms (three derived groups plus $\alpha$). The form of the equation we are looking for can be expressed as
\begin{equation}
\Theta_{\rm NDR} = \mathcal{F}\left(\Delta\nu \Delta t,\,  \nu_p \Delta t \, , \,  \frac{f \, \Delta t}{h \, DM},\,  \alpha\right),
\label{eq:dimensionless-equation}
\end{equation}

where the threshold can be defined by our preference, for example, as being smaller than 0 for non-repeaters and larger than 0 for repeaters. This equation avoids degeneracy, aligns with astrophysical scaling laws, and preserves physical clarity.

\subsection{Probabilistic Model}
\label{sec:Probabilistic}
Since this is a classification problem, we expect the equation $\Theta_{\rm NDR}$ to represent the distributions of the two classes. Thereby allowing us to compute the probability of an event being classified as either a repeater or a non-repeater by the classical Sigmoid function after normalizing to $\bar{\Theta}_{\rm NDR}$
\begin{equation}
	\sigma(\bar{\Theta}_{\rm NDR}) = \frac{1}{1 + e^{-\bar{\Theta}_{\rm NDR}}}.
\label{eq:sigmoid}
\end{equation}
The output range of $\sigma(\bar{\Theta}_{\rm NDR})$ is $(0,1)$, representing the probability that the classification belongs to the positive class (repeater), denoted as $p$. The probability of the negative class (non-repeater) is then $1 - p$. In a typical binary classification problem, if $p \geq 0.5$ (i.e., $ \bar{\Theta}_{\rm NDR} \geq 0$), the instance is predicted as the positive class (repeater). If $p < 0.5$ (i.e., $\bar{\Theta}_{\rm NDR} < 0$), it is predicted as the negative class (non-repeater).

We employ PySR for symbolic regression and restrict the operator set to $\{+, -, \times, \div, \text{pow}\}$ for simplicity. Addition and subtraction represent the linear superposition of physical quantities, while multiplication and division capture relationships such as proportional scaling, aligning with the mathematical structures of classical physical laws. The \text{pow} operator is introduced to accommodate the nonlinear relations, and the potential approximations involving $f$ and $DM$. These two parameters may require integration, which our equation does not explicitly incorporate, we approximate it using an exponential function.


To derive the equation, we minimize the cross-entropy loss
\begin{equation}
    \begin{aligned}
        L(\hat{\Theta}, \bar{\Theta}_{\rm NDR}) = 
        & - \hat{\Theta} \log \sigma(\bar{\Theta}_{\rm NDR}) \\
        & - (1 - \hat{\Theta}) \log (1 - \sigma(\bar{\Theta}_{\rm NDR})).
    \end{aligned}
\end{equation}
where $\hat{\Theta}$ is the ground-truth labels, taken as the catalog's classifications. In information theory, entropy quantifies the uncertainty in a system, and cross-entropy measures how much extra information is needed when using an estimated distribution instead of the true one. Therefore, minimizing cross-entropy loss decreases the discrepancy between the equation’s predictions and the catalog’s classifications, guiding the model towards more accurate predictions with less informational inefficiency.

Notably, we separated the equation of $\Theta_{\rm NDR}$ from the nonlinear probability equation and the loss function, rather than directly optimizing $\Theta_{\rm NDR}$ to output classification probabilities. This procedure allows $\Theta_{\rm NDR}$ retains a simple, physically interpretable form while reducing operator complexity and constraining the search space of the genetic algorithm.

All found candidate equations strictly follow the constraint of dimensional homogeneity since the input dimensionless groups have eliminated the influence of the original dimensions, and the operator combination process always maintains dimensional balance. Since we are dealing with a classification task, the equation output must be a dimensionless quantity, which is naturally ensured by the direct combination of these dimensionless groups.

Considering the generalizability of the equation, we prioritize simplicity in its derivation rather than solely maximizing classification accuracy on Catalog 1. This is because the data itself contains uncertainties, and there is no prior conclusion that the two classifications can be perfectly distinguished using observational parameters. As a result, the accuracy of the equation is likely to have an upper limit. Over-optimizing for accuracy may lead to overfitting, where the equation captures specific features of certain bursts in Catalog 1 rather than the underlying general properties. Therefore, we set a recall and F2-score threshold of 80\% for the initial filtering and, under the primary constraint of simplicity, aim for higher accuracy.

Technically, under the PySR framework, the ``best'' scoring rule governs the evolutionary search by explicitly balancing mathematical simplicity and predictive accuracy. This dual-objective optimization penalizes excessive complexity through operator-specific weights (e.g., imposing higher costs for exponentiation than addition), thereby enforcing Occam's Razor to mitigate overfitting while retaining physically meaningful structures.

\section{Results}
\label{sec:results}
\subsection{Power-Law Model: Results}
\begin{figure*}[htbp!]
    \centering
    \includegraphics[width=0.353\linewidth]{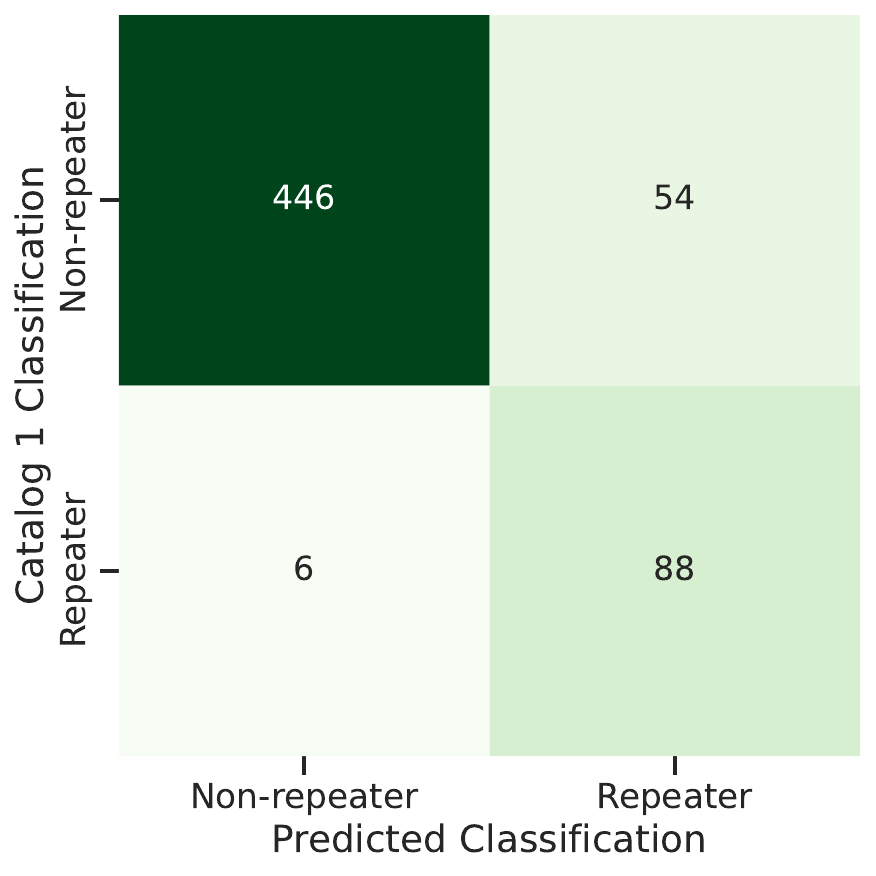}
    \includegraphics[width=0.446\linewidth]{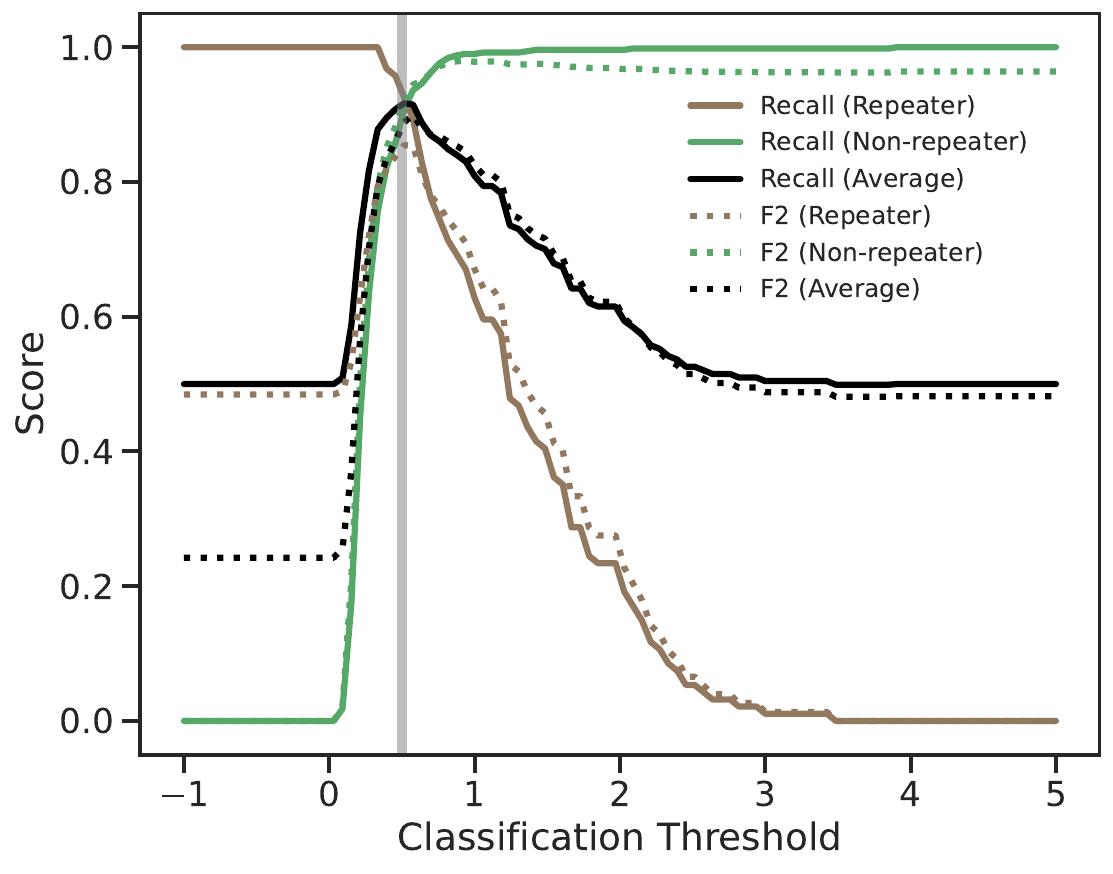}
    \caption{Classification of Catalog 1 by Power-law Multiplication. \textbf{Left:} Confusion matrix by applying Eq. \ref{eq:powerlaw} with a threshold of 0. \textbf{Right:} Recall and F2 score by different threshold. Green represents non-repeaters, brown represents repeaters, and the grey vertical line indicates the threshold at 0.5.}
    \label{fig:cat1-pl}
\end{figure*}


To fit and to evaluate the effectiveness of the power-law multiplication method in distinguishing between repeating and non-repeating FRBs, we applied a linear regression model using the scikit-learn package \cite{scikit-learn} to Catalog 1 \cite{chime/frbcollaboration2021FirstCHIMEFRB}. The regression analysis aimed to determine the coefficients $\beta_0, \beta_1, \dots, \beta_6$ as shown in Equation (\ref{eq:linef}). 

To mitigate variability in the equation's performance, we performed 100 randomized train-test splits of the dataset. The final coefficients are selected based on the iteration achieving the highest F2 score across all Catalog 1 bursts. Meanwhile, we should note that the accuracy of these coefficients is not guaranteed due to the impact of the training set, which includes possible mislabeled data, simulated repeater samples, and data with varying distributions.

Despite these constraints, the optimized equation achieves strong classification performance, with an F2-score of $0.88$ and a recall of $0.91$ at the classification threshold of $0.5$. The equation writes

\begin{equation}
\begin{aligned}
  \Theta_{\rm PL} = \, & 0.45 \, e^{0.013 \alpha} 
  \left( \frac{f}{1~\rm{Jy}} \right)^{-0.20} 
  \left( \frac{DM}{100~\rm{pc~cm^{-3}}} \right)^{-0.34} \\
  &\left( \frac{\Delta \nu}{200 ~\rm{Hz}} \right)^{-0.44}  
  \left( \frac{\Delta t}{1 ~\rm{ms}} \right)^{0.31}  
  \left( \frac{\nu_p}{500 ~\rm{Hz}} \right)^{0.75}.
  \label{eq:powerlaw}
\end{aligned}
\end{equation}

The bursts with $\Theta_{\rm PL} \leq 0.5$ are classified as non-repeaters, while $\Theta_{\rm PL} > 0.5$ are classified as repeaters. Figure \ref{fig:cat1-pl} shows its classification results. 

\subsection{Power-Law Model: Physical Interpretation}

From the absolute value of the exponent in the fitted equation, we can determine the sensitivity of each parameter to classification, i.e., how a change of one order of magnitude affects the result. It can be observed that $\nu_p$ has the highest sensitivity, while $f$ has the lowest. For the $\alpha$ term, we have $e^{0.013 \alpha} = 10^{0.0056 \alpha}$, indicating that its sensitivity on the order of magnitude is $ 0.0056 \alpha $, which increases as $\alpha$ grows.

If we consider the impact degree of each parameter on the result, we must also account for the order of magnitude of the parameter's value range, as shown in Figure~\ref{fig:para_dis}. This gives the order-of-magnitude changes for each parameter's entire range: $\alpha$: 0.61, $f$: -0.48, $DM$: -0.71, $\Delta \nu$: -0.429, $\Delta t$: 0.75 and $\nu_p$: 0.23, which are largely consistent with the rankings from the parameter selection of neural networks shown in Figure~\ref{fig:select_feature2}.

This equation not only captures the tendency of each parameter toward classification but also reflects the relationships among parameters. For example, considering $\Delta \nu$ and $\nu_p$, Figure~\ref{fig:para_dis} shows that repeaters tend to have higher $\nu_p$ and lower $\Delta \nu$. Correspondingly, the exponents in the equation are $0.75$ and $-0.44$, which align with this tendency. Furthermore, this is consistent with the relationship $\Delta \nu \approx \nu_p^2$ shown in Figure~\ref{fig:vp-dv-relation}, proved by substituting this relation into the equation, the $\Delta \nu$ and $\nu_p$ terms nearly cancel each other out ($-0.88$ vs. $0.75$). 

From the equation, repeaters have steep spectral indices ($\alpha$), higher peak frequencies ($\nu_p$), narrower bandwidths ($\Delta \nu$), longer durations ($\Delta t$), and a strong correlation between peak frequency ($\nu_p$) and bandwidth ($\Delta \nu$). Any proposed competitive FRB radiation mechanism should explain these differences.

The dispersion measure ($DM$) appears to indicate that repeaters are more localized, a possible interpretation is that nearby ones have lower luminosities and the repetition rate is luminosity dependent. Usually it takes longer for high-luminosity bursts to repeat. The far-away non-repeaters could be simply repeaters that just take longer to repeat. We also need to be cautious about the limited sample size that each burst from a repeating source has different radiation features but the same $DM$, 94 bursts actually originate from only 18 sources, meaning there are effectively only 18 independent $DM$ values. Insufficient data may cause inaccurate inferences.

Flux ($f$) is the least sensitive parameter, repeaters and non-repeaters have similar energy magnitudes. This suggests they might share a common engine source, such as magnetic energy from a compact star, while differing in their energy release mechanisms.

This empirical equation relies on a linear combination of log-transformed parameters (via power-law scaling) to classify bursts. While it reflects the sensitivity and impact of each parameter on distinguishing populations, this approach has inherent limitations in capturing nonlinear or complex relationships, which we will address by the next approach.

\subsection{NDR: Results}

We have constructed the four dimensionless groups derived from Buckingham's $\pi$ theorem, transforming observed parameters into dimensionless physical forms.

\begin{figure*}[htbp!]
    \centering
    \includegraphics[width=0.353\linewidth]{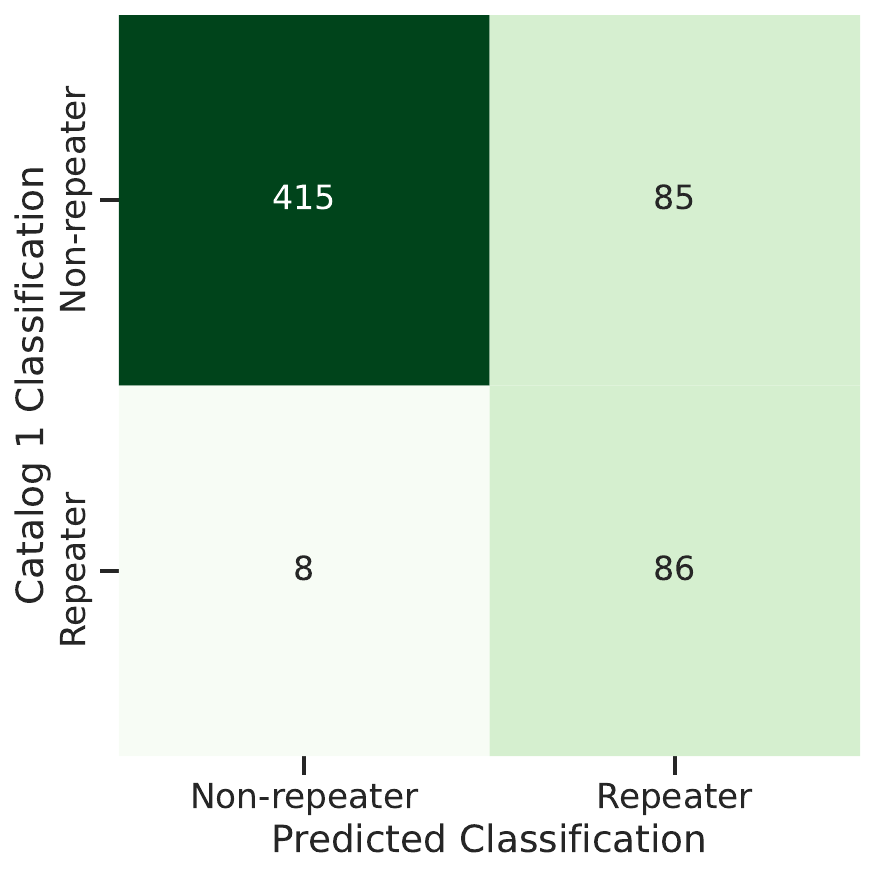}
    \includegraphics[width=0.446\linewidth]{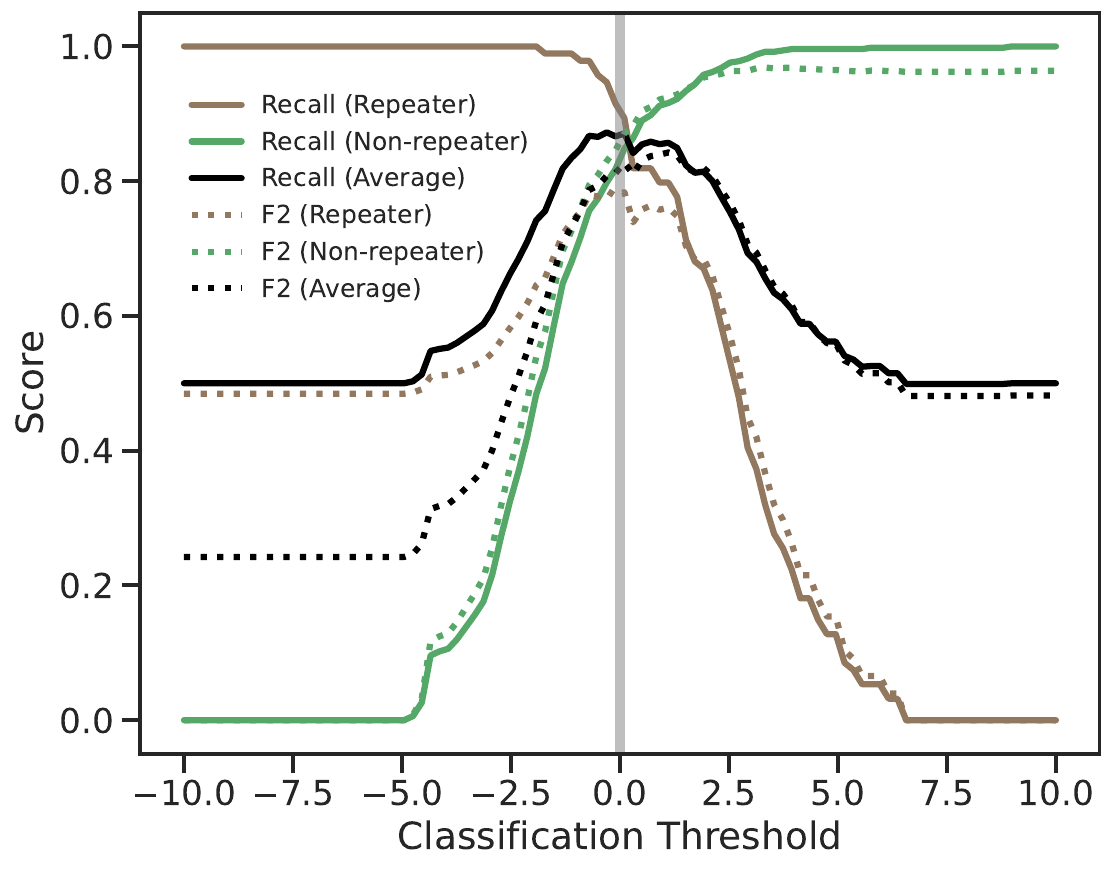}
    \caption{Classification of Catalog 1 FRBs by NDR. \textbf{Left:} Confusion matrix by appling equation \ref{eq:ndr2} and threshold is 0. \textbf{Right:} Recall and F2 score of different threshold. Green represents non-repeaters, brown represents repeaters, and the grey vertical line indicates the threshold at 0.}
    \label{fig:cat2-bdr}
\end{figure*}

\begin{figure*}
    \centering
    \includegraphics[width=0.85\linewidth]{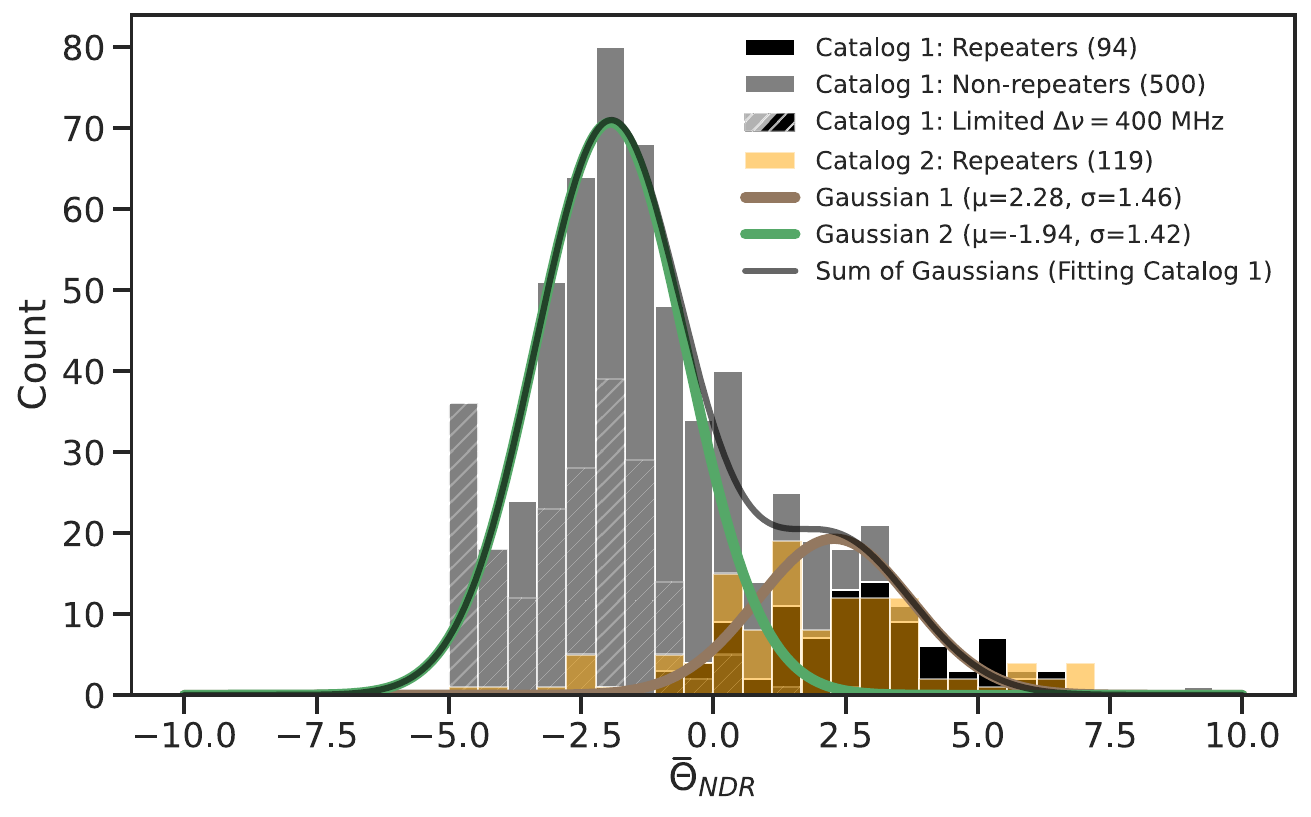}
    \caption{ Distribution of $\bar{\Theta}_{\text{NDR}}$ values for different FRB populations. The stacked histogram shows Catalog 1 repeaters (black) and non-repeaters (grey), with hatched regions indicating sources constrained by the bandwidth limit ($\Delta \nu = 400$ MHz). The orange histogram represents Catalog 2 repeaters. The solid color curves show the individual Gaussian components of the bimodal fit (Brown: Gaussian 1 enclosing repeaters; Green: Gaussian 2 enclosing non-repeaters), while the solid black line shows their sum. This bimodal distribution suggests two distinct populations within the FRB sample.}
    \label{fig:ThetaNDR}
\end{figure*}

Due to the stochastic nature of the genetic algorithm in PySR, the optimal equation obtained varies in each run. We performed 10,000 fittings, generating 10,000 candidate equations, and then ranked them by F2-score and recall (many equations output identical classification scores). We observed that the equations with recall and F2-score over $0.8$ mostly do not include the $\pi_3$ term, while some combinations of $\pi_1$ and $\pi_2$ frequently appear. Finally, we selected the a simple equation with F2-score $0.82$ and recall $0.87$, as   
\begin{equation}
\Theta_{\rm NDR} =  (\pi_2 - \pi_1)^{0.3} + \frac{\pi_2}{0.3 \pi_1} + \pi_4
\label{eq:ndr1}
\end{equation}
or in the form of the observables 
\begin{equation}
\Theta_{\rm NDR} = (\nu_p - \Delta \nu)^{0.3} \Delta t^{0.3} + \frac{\nu_p}{0.3 \Delta \nu} + \alpha
\label{eq:ndr2}
\end{equation}
As expected, $\Theta_{\rm NDR}$ can be well fitted by the superposition of two Gaussian distributions, see Figure \ref{fig:ThetaNDR}, which is the most important result of this article. The Gaussian distribution with a lower mean contains most of the non-repeaters (gray histogram bars), while the Gaussian distribution with a higher mean primarily includes the repeaters (black histogram bars). A small number of non-repeaters also fall within the Gaussian distribution with the higher mean, these non-repeaters might be repeaters for which the repeating phenomenon has not yet been observed. The limitations from observation are evident in the distribution. For instance, in Figure \ref{fig:ThetaNDR}, a bin of exceptional high count appears on the far left. This occurs because CHIME's bandwidth is restricted to 400 MHz, causing tens of FRBs with an actual bandwidth exceeding 400 MHz to be accumulated in this bin.

To compute the probability, we first normalize the equation to center it around 0 and scale it to make its variance in the order of 1, then apply the Sigmoid function of equation \ref{eq:sigmoid}. Our regression method optimizes this normalization as  
\begin{equation}  
\bar{\Theta}_{\rm NDR} = \frac{\Theta_{\rm NDR} - 77}{15.4}.  
\label{eq:ndr3}  
\end{equation}  
Here, the constant $77$ shifts the distribution, while the factor $1/15.4$ scales it. This normalization does not affect the classification capability of the equation. The bursts with $\bar{\Theta}_{\rm NDR} \leq 0$ are classified as non-repeaters, while  $\bar{\Theta}_{\rm NDR} > 0$ are classified as repeaters.


\subsection{NDR: Physical Interpretation}

Equation $\ref{eq:ndr2}$ shows that classification is related to $\nu_p$, $\Delta_\nu$, $\delta t$, and $\alpha$, but independent of $DM$ and $f$. Since $DM$ serves as a proxy for redshift $z$, it implies the two FRB classes do not exhibit significant differences in their cosmological distributions. Given that currently observed FRBs are within $z < 1$, future observations at higher redshifts may clarify whether this trend persists or diverges. The flux ($f$), which correlates with intrinsic luminosity and distance, further implies that the two classes lack statistically significant differences in luminosity. If their redshift distributions are similar from above discussion, this suggests that their energy scales and emission mechanisms may originate from progenitors with comparable energy reservoirs, such as magnetars or neutron stars. 

The parameters $\nu_p$, $\Delta_\nu$, and $\alpha$ describe the spectral characteristics, while $\Delta t$ represents the burst duration. These quantities are related to radiation mechanisms, the physical scale of the emission process, and the duration of active central engine. From the equation, its value increases as the difference ($\nu_p - \Delta_\nu$) and the ratio ($\nu_p / \Delta_\nu$) between the spectral peak frequency and bandwidth, as well as the spectral index ($\alpha$) and the duration ($\delta t$), increase. This suggests that repeaters tend to exhibit higher-frequency, narrower, steeper spectra and longer durations compared to non-repeaters. In contrast, non-repeaters tend to exhibit broader spectral widths, flatter spectra, and shorter durations. This conclusion aligns with the previous discussion based on equation $\ref{eq:powerlaw}$.

It is interesting to notice $DM$, representing redshift, is absent in the equation \ref{eq:ndr2} while having high rankings during the feature selection. This discrepancy may arise from differences between traditional neural network and symbolic equation. 

A neural network, with its many layers and interconnected nodes, acts as a complex formula of thousands or even millions of parameters. It learns complex, nonlinear relationships between the six input parameters (including $DM$ and $f$) and the FRB classification. The SHAP values for feature selection provide relative importance within the network's learned representation, not necessarily the absolute importance in the underlying physics. The neural network needs $DM$ because it's a proxy for redshift, which affects other observables and highly involved in computation. Moreover, our data contain uncertainties and observational limitations, which further prevent the neural network from clearly evaluating the contribution of the highly entangled $DM$ and to eliminate it.

Our derived equation, however, identifies that only peak frequency ($\nu_p$), bandwidth ($\Delta \nu$), spectral index ($\alpha$), and duration ($\Delta t$) are fundamental parameters. They combine in a redshift-independent manner due to opposing scaling from cosmic expansion: $\nu_p$ and $\Delta \nu$ scale as $(1+z)^{-1}$, $\Delta t$ scales as $(1+z)$, and $\alpha$ remains unaffected. The terms $(\nu_p - \Delta \nu) \Delta t$ and $\nu_p / \Delta \nu$ mathematically cancel $(1+z)$ dependencies. Consequently, $DM$, which encodes redshift information, becomes unnecessary for this simple equation. As more samples are included in the future observations, the $DM$ term may enter into more detailed equations.

This highlights the difference between neural networks and symbolic regression. A complex neural network cannot eliminate quantities that are indirectly involved in computations but ultimately irrelevant to the final result. In contrast, a concise mathematical equation can eliminate such dependencies.

\subsection{Application on 2nd Catalog}

\label{sec:application2nd}




In 2023, CHIME released its second repeater-only FRB catalog (Catalog 2), see Section \ref{sec:data} for details. We applied the equations derived using the power-law multiplication and NDR method to the 119 sub-bursts of this new catalog. 

\begin{figure*}[t]
    \centering
    \includegraphics[width=0.35\linewidth]{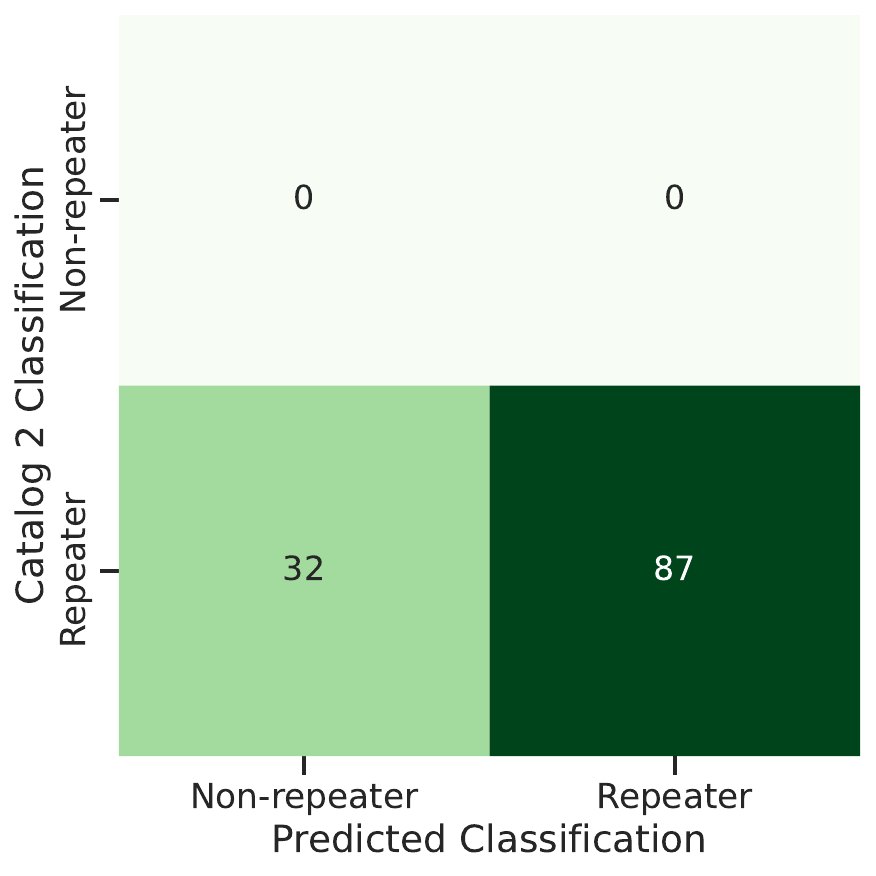}
    \includegraphics[width=0.446\linewidth]{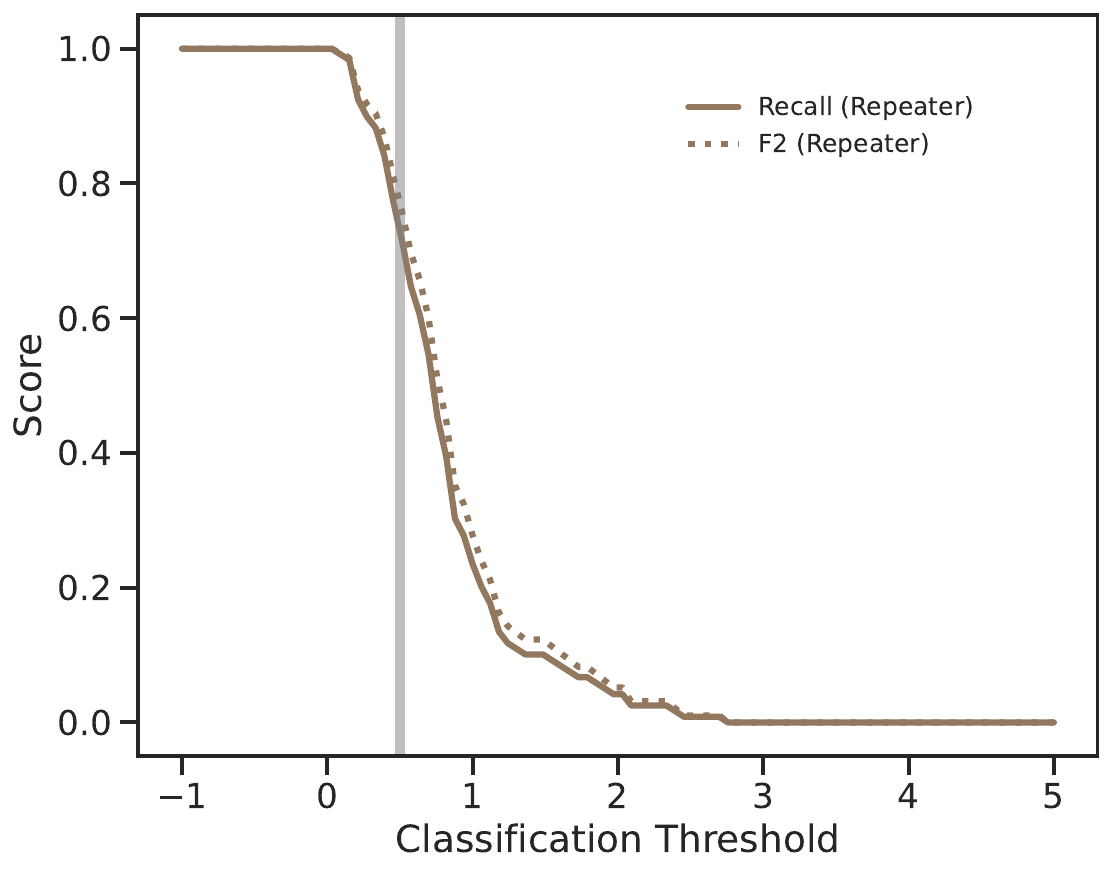}
     \includegraphics[width=0.35\linewidth]{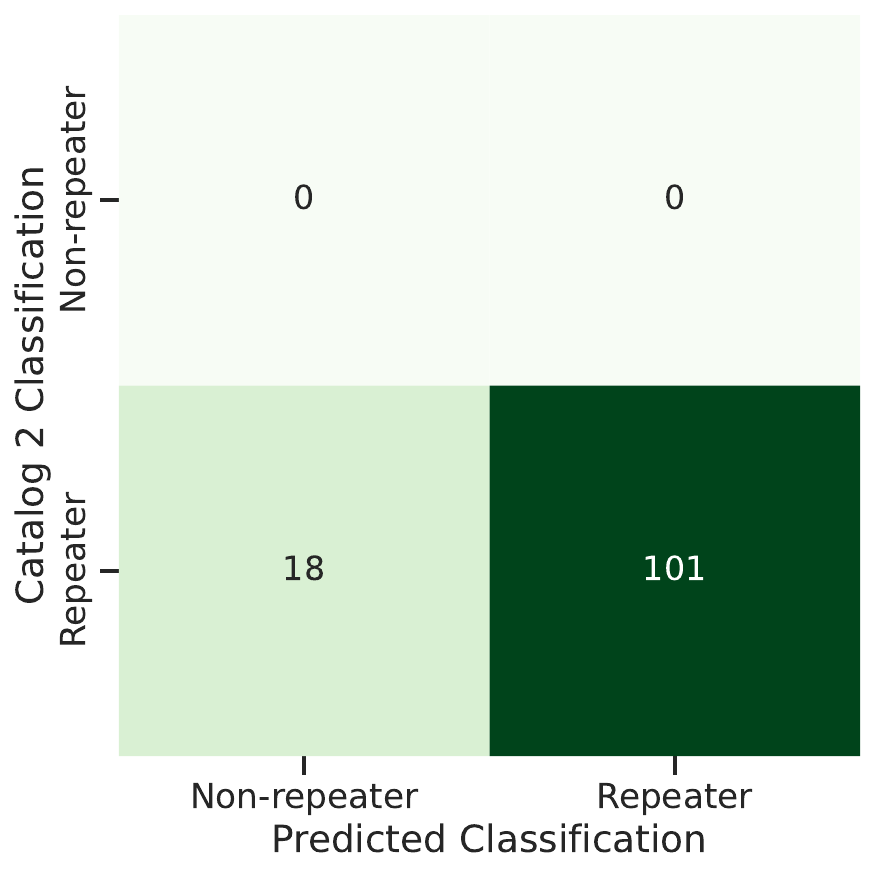}
    \includegraphics[width=0.446\linewidth]{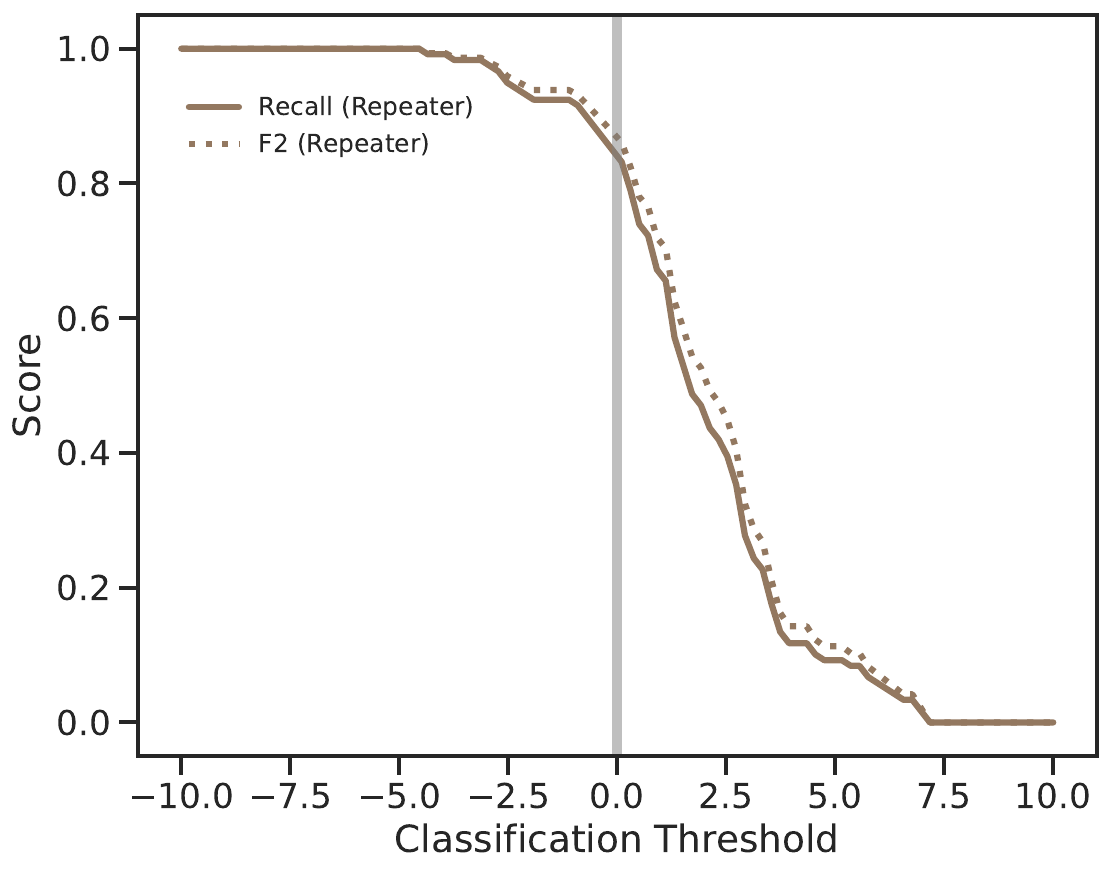}
\caption{Confusion matrices, recall, and F2 scores for power-law multiplication and NDR methods for catalog 2 classification, which include repeaters only. 
\textbf{Upper-Left:} Power-law multiplication confusion matrix using Equation \ref{eq:powerlaw} (threshold = 0.5).  
\textbf{Upper-Right:} Recall and F2 score for power-law multiplication at varying thresholds; green (non-repeaters), brown (repeaters), grey vertical line (threshold = 0.5).  
\textbf{Lower-Left:} NDR confusion matrix using Equation \ref{eq:ndr2} (threshold = 0).  
\textbf{Lower-Right:} Recall and F2 score for NDR at varying thresholds; green (non-repeaters), brown (repeaters), grey vertical line (threshold = 0).}
    \label{fig:cat2}
\end{figure*}

The power-law multiplication, Equation \ref{eq:powerlaw}, infers 87 repeaters, achieving a recall\footnote{Since catalog 2 contains only repeaters, recall is equivalent to accuracy.} of 0.73, which is lower than its recall of 0.91 on catalog 1. The NDR method, Equation \ref{eq:ndr2}, achieves a recall of 0.85, which is close to its recall of 0.87 on catalog 1. See Figure \ref{fig:cat2} for the results. It is remarkable that Equation \ref{eq:ndr2}, a simple equation composed of only four observational parameters, achieves such a high accuracy, especially considering that the data itself is constrained and contains uncertainties, as well as the catalog itself could be misclassified. 


The equation derived from power-law multiplication, which incorporates all six parameters, performed exceptionally well on Catalog 1 but exhibits a noticeable decline in performance on Catalog 2. In contrast, the equation derived from NDR demonstrates greater stability. This suggests that the power-law multiplication equation may be prone to overfitting on Catalog 1, reducing its generalizability. Such an issue is commonly observed when the dataset is small, particularly with only 18 independent $DM$ measured, and when the data is imprecise due to instrumental limitations and measurement errors.

In contrast, the simpler equation derived from NDR appears to be more aligned with the fundamental physics underlying classification. Additionally, it partitions the data into two distributions. When we plot the Catalog 2 results on Figure \ref{fig:ThetaNDR}, the data points fall within the distribution of repeaters of Catalog 1, further suggesting that this equation may have some capacity to reflect two distinct physical processes. 

As a robustness check, Appendix~\ref{sec:Robustness} shows that our NDR equation recovers 5/6 Catalog-2 repeaters previously labeled non-repeating in Catalog 1. In Appendix~\ref{sec:Candidates}, we present a list of repeater candidates derived from the NDR equation and cross-validate it against the results of \cite{2022luo,2022zhuge_nosup}.

From a statistical perspective, the overall distribution of categories carries more significance. Even if some data labels are incorrect, a meaningful distribution can still emerge as long as the majority are correctly classified. This distribution serves as an important reference for assessing whether unsure bursts, those repeaters classified as non-repeaters due to inadequate observational time for confirmation, are classified correctly. We also need to note that the two distributions still overlap, indicating that the equation cannot provide a definitive classification for samples with derived values near zero. This situation is similar to GRB classification using $T_{90}$, where two slightly overlapping distributions correspond to different GRB origins. However, GRBs near the $2$ s threshold cannot be perfectly distinguished using $T_{90}$ alone and require additional criteria, such as the presence of a supernova or kilonova. If FRBs indeed have two or more distinct origins, future observations in other wavebands may reveal additional distinguishing evidence.

\section{Conclusion and Discussion}
\label{sec:conclusion}

In this work, we demonstrated how human expertise and machine learning can synergize to derive interpretable phenomenological equations, and gave the example of classifying FRBs based solely on physical observables from CHIME telescope.

First, the neural network helped identify the top six observational parameters, emerged from 48 observational parameters, including spectral index ($\alpha$), sub-burst duration ($\Delta t$), dispersion measure ($\text{DM}$), flux density ($f$), frequency bandwidth ($\Delta \nu$), and peak frequency ($\nu_p$). These parameters provide a comprehensive description of FRB observations, each of them plays a unique role that cannot be substituted by another, suggesting that the mechanisms behind repeating and non-repeating FRBs could be distinct.

Then we search for the equations using these six parameters by two methods. For the equation \ref{eq:powerlaw} made of power-laws multiplication, we achieved F2 score of $0.88$ and a recall rate of $0.91$. Its power-law indices reveal that repeating bursts tend to exhibit steeper spectral indices ($\alpha$), higher peak frequencies ($\nu_p$), narrower bandwidths ($\Delta \nu$), and longer durations ($\Delta t$). In contrast, non-repeating bursts showed the opposite spectral trends.

To further investigate the underlying physics and the nonlinear relationships between parameters. We propose a second approach, neural dimensional regression, which combines human prior knowledge with machine learning. Through an initial dimensionless processing, this method reduces the number of parameters and significantly decreases the computational complexity, it also ensures the dimension consistency that the traditional symbolic regression may find problem. Based on Buckingham's $\pi$-theorem, we selected high-impact, low-nonlinearity parameters as repeating variables and constructed four dimensionless groups: $\pi_1 = \Delta \nu \Delta t$, $\pi_2 = \nu_p \Delta t$, $\pi_3 = f \Delta t / (h \text{DM})$, and $\pi_4 = \alpha$. Then by the symbolic regression (using the PySR library), we obtained a simple but highly performing equation \ref{eq:ndr2} with an F2 score of $0.82$ and a recall rate of $0.87$. Critically, this NDR equation partitioned FRBs into two distinct Gaussian distributions (Figure \ref{fig:ThetaNDR}), similar to GRB's $T_{90}$ based classification, suggesting two origins. 

We then apply the equations on the second CHIME catalog which only contains repeaters. The NDR-based equation achieved a recall of $0.85$, maintaining stability across catalogs, indicating that it is less influenced by dataset variations and may capture underlying physical principles. The Power-law multiplication equation drops to $0.73$, suggesting it is mildly over-fitted for the first catalog. The classification output of the second catalog using the NDR equation falls within the repeater distribution of the first catalog, further validating the equation's physical significance and confirming the two classes.


We acknowledge potential limitations. For the observations, the classification is derived from CHIME data within the 400–800 MHz band. Its applicability to data from instruments observing at different frequencies requires verification. Furthermore, sample sizes for some parameters, particularly the DM values for repeaters, remain limited as fewer than 20. For the results, this work concludes that there are two classes of FRBs based on the currently available limited data. However, it does not determine whether these classes arise from distinct progenitor systems, from a single progenitor with different emission mechanisms, or from a common mechanism operating under varying physical conditions, such as active and inactive phases. These possibilities will be investigated in future work.

In conclusion, this work exemplifies how human-driven physical reasoning, dimensional homogeneity, nonlinear interaction analysis, and Occam’s Razor, can steer symbolic regression toward robust, interpretable laws. Bridging machine learning’s pattern recognition with domain knowledge helps advance the search for fundamental physical descriptions in complex astrophysical datasets. Future work incorporating polarimetry, multi-wavelength data, and more samples could further refine these laws, probing the progenitor systems behind FRB diversity.

\Acknowledgements{We thank the CHIME/FRB Collaboration for providing access to publicly released data products. We also thank Professor Di Li at Tsinghua University for helpful discussions that improved this work. This work was supported in part by the National Natural Science Foundation of China (Grant Nos. 12475057 and 12303079).
}%


\InterestConflict{The authors declare that they have no conflict of interest.}

\bibliographystyle{BibStyle}
\bibliography{main}

\providecommand{\href}[2]{#2}\begingroup\raggedright\begin{thebibliography}{10}

\bibitem{husserl2001phenomenology}
E.~Husserl, T.~Klein, and W.~Pohl, \emph{Phenomenology and the Foundations of the Sciences}, Collected works (Husserl, Edmund). M. Nijhoff, 2001.

\bibitem{moran2002introduction}
D.~Moran, \emph{Introduction to phenomenology}. routledge, 2002.

\bibitem{2015Sci...349..255J}
M.~I. {Jordan} and T.~M. {Mitchell}, \href{https://doi.org/10.1126/science.aaa8415}{Science {\bfseries 349}, 255 (2015)}.

\bibitem{2019RvMP...91d5002C}
G.~{Carleo}, I.~{Cirac}, K.~{Cranmer}, L.~{Daudet}, M.~{Schuld}, N.~{Tishby}, L.~{Vogt-Maranto}, and L.~{Zdeborov{\'a}}, \href{https://doi.org/10.1103/RevModPhys.91.045002}{Reviews of Modern Physics {\bfseries 91}, 045002 (2019)}, arXiv: \href{https://arxiv.org/abs/1903.10563}{{1903.10563}}.

\bibitem{makke2024interpretable}
N.~Makke and S.~Chawla, {Artificial Intelligence Review {\bfseries 57}, 2 (2024)}.

\bibitem{fortin2012deap}
F.-A. Fortin, F.-M. De~Rainville, M.-A.~G. Gardner, M.~Parizeau, and C.~Gagn{\'e}, {The Journal of Machine Learning Research {\bfseries 13}, 2171 (2012)}.

\bibitem{2023arXiv230501582C}
M.~{Cranmer}, \href{https://doi.org/10.48550/arXiv.2305.01582}{arXiv e-prints arXiv:2305.01582 (2023)}, arXiv: \href{https://arxiv.org/abs/2305.01582}{{2305.01582}}.

\bibitem{2024arXiv240419756L}
Z.~{Liu}, Y.~{Wang}, S.~{Vaidya}, F.~{Ruehle}, J.~{Halverson}, M.~{Solja{\v{c}}i{\'c}}, T.~Y. {Hou}, and M.~{Tegmark}, \href{https://doi.org/10.48550/arXiv.2404.19756}{arXiv e-prints arXiv:2404.19756 (2024)}, arXiv: \href{https://arxiv.org/abs/2404.19756}{{2404.19756}}.

\bibitem{2023NatCo..14.1777C}
C.~{Cornelio}, S.~{Dash}, V.~{Austel}, T.~R. {Josephson}, J.~{Goncalves}, K.~L. {Clarkson}, N.~{Megiddo}, B.~{El Khadir}, and L.~{Horesh}, \href{https://doi.org/10.1038/s41467-023-37236-y}{Nature Communications {\bfseries 14}, 1777 (2023)}.

\bibitem{2023MLS&T...4d5002L}
P.~{Lemos}, N.~{Jeffrey}, M.~{Cranmer}, S.~{Ho}, and P.~{Battaglia}, \href{https://doi.org/10.1088/2632-2153/acfa63}{Machine Learning: Science and Technology {\bfseries 4}, 045002 (2023)}, arXiv: \href{https://arxiv.org/abs/2202.02306}{{2202.02306}}.

\bibitem{2021PhRvB.104w5111M}
C.~{Miles}, M.~R. {Carbone}, E.~J. {Sturm}, D.~{Lu}, A.~{Weichselbaum}, K.~{Barros}, and R.~M. {Konik}, \href{https://doi.org/10.1103/PhysRevB.104.235111}{\prb {\bfseries 104}, 235111 (2021)}, arXiv: \href{https://arxiv.org/abs/2107.08013}{{2107.08013}}.

\bibitem{2024CSF...18815538D}
M.~{De Florio}, I.~G. {Kevrekidis}, and G.~E. {Karniadakis}, \href{https://doi.org/10.1016/j.chaos.2024.115538}{Chaos Solitons and Fractals {\bfseries 188}, 115538 (2024)}, arXiv: \href{https://arxiv.org/abs/2312.14237}{{2312.14237}}.

\bibitem{2022NatRP...4..399K}
G.~{Karagiorgi}, G.~{Kasieczka}, S.~{Kravitz}, B.~{Nachman}, and D.~{Shih}, \href{https://doi.org/10.1038/s42254-022-00455-1}{Nature Reviews Physics {\bfseries 4}, 399 (2022)}, arXiv: \href{https://arxiv.org/abs/2112.03769}{{2112.03769}}.

\bibitem{angelis2023artificial}
D.~Angelis, F.~Sofos, and T.~E. Karakasidis, {Archives of Computational Methods in Engineering {\bfseries 30}, 3845 (2023)}.

\bibitem{lorimer2007bright}
D.~R. Lorimer, M.~Bailes, M.~A. McLaughlin, D.~J. Narkevic, and F.~Crawford, {Science {\bfseries 318}, 777 (2007)}.

\bibitem{thornton2013population}
D.~e. Thornton, B.~Stappers, M.~Bailes, B.~Barsdell, S.~Bates, N.~Bhat, M.~Burgay, S.~Burke-Spolaor, D.~Champion, P.~Coster, et~al., {Science {\bfseries 341}, 53 (2013)}.

\bibitem{2023RvMP...95c5005Z}
B.~{Zhang}, \href{https://doi.org/10.1103/RevModPhys.95.035005}{Reviews of Modern Physics {\bfseries 95}, 035005 (2023)}, arXiv: \href{https://arxiv.org/abs/2212.03972}{{2212.03972}}.

\bibitem{2018Katz_FRB}
J.~I. {Katz}, \href{https://doi.org/10.1016/j.ppnp.2018.07.001}{Progress in Particle and Nuclear Physics {\bfseries 103}, 1 (2018)}, arXiv: \href{https://arxiv.org/abs/1804.09092}{{1804.09092}}.

\bibitem{Popov_2018}
S.~B. Popov, K.~A. Postnov, and M.~S. Pshirkov, \href{https://doi.org/10.3367/ufne.2018.03.038313}{Physics-Uspekhi {\bfseries 61}, 965 (2018)}.

\bibitem{2019Cordes_FRB}
J.~M. {Cordes} and S.~{Chatterjee}, \href{https://doi.org/10.1146/annurev-astro-091918-104501}{\araa {\bfseries 57}, 417 (2019)}, arXiv: \href{https://arxiv.org/abs/1906.05878}{{1906.05878}}.

\bibitem{Petroff2019}
E.~Petroff, J.~W.~T. Hessels, and D.~R. Lorimer, \href{https://doi.org/10.1007/s00159-019-0116-6}{The Astronomy and Astrophysics Review {\bfseries 27}, 4 (2019)}.

\bibitem{2007Lorimer}
D.~R. {Lorimer}, M.~{Bailes}, M.~A. {McLaughlin}, D.~J. {Narkevic}, and F.~{Crawford}, \href{https://doi.org/10.1126/science.1147532}{Science {\bfseries 318}, 777 (2007)}, arXiv: \href{https://arxiv.org/abs/0709.4301}{{0709.4301}}.

\bibitem{ai2021true}
S.~Ai, H.~Gao, and B.~Zhang, {The Astrophysical Journal Letters {\bfseries 906}, L5 (2021)}.

\bibitem{chime/frbcollaboration2021FirstCHIMEFRB}
M.~{Amiri}, B.~C. {Andersen}, K.~{Bandura}, S.~{Berger}, M.~{Bhardwaj}, M.~M. {Boyce}, P.~J. {Boyle}, C.~{Brar}, D.~{Breitman}, T.~{Cassanelli}, P.~{Chawla}, T.~{Chen}, J.~F. {Cliche}, A.~{Cook}, D.~{Cubranic}, A.~P. {Curtin}, M.~{Deng}, M.~{Dobbs}, F.~{(Adam) Dong}, G.~{Eadie}, M.~{Fandino}, E.~{Fonseca}, B.~M. {Gaensler}, U.~{Giri}, D.~C. {Good}, M.~{Halpern}, A.~S. {Hill}, G.~{Hinshaw}, A.~{Josephy}, J.~F. {Kaczmarek}, Z.~{Kader}, J.~W. {Kania}, V.~M. {Kaspi}, T.~L. {Landecker}, D.~{Lang}, C.~{Leung}, D.~{Li}, H.-H. {Lin}, K.~W. {Masui}, R.~{McKinven}, J.~{Mena-Parra}, M.~{Merryfield}, B.~W. {Meyers}, D.~{Michilli}, N.~{Milutinovic}, A.~{Mirhosseini}, M.~{M{\"u}nchmeyer}, A.~{Naidu}, L.~{Newburgh}, C.~{Ng}, C.~{Patel}, U.-L. {Pen}, E.~{Petroff}, T.~{Pinsonneault-Marotte}, Z.~{Pleunis}, M.~{Rafiei-Ravandi}, M.~{Rahman}, S.~M. {Ransom}, A.~{Renard}, P.~{Sanghavi}, P.~{Scholz}, J.~R. {Shaw}, K.~{Shin}, S.~R. {Siegel}, A.~E. {Sikora}, S.~{Singh}, K.~M. {Smith}, I.~{Stairs}, C.~M. {Tan}, S.~P. {Tendulkar},
  K.~{Vanderlinde}, H.~{Wang}, D.~{Wulf}, A.~V. {Zwaniga}, and {CHIME/FRB Collaboration}, \href{https://doi.org/10.3847/1538-4365/ac33ab}{\apjs {\bfseries 257}, 59 (2021)}.

\bibitem{2022zhuge_nosup}
J.-M. Zhu-Ge, J.-W. Luo, and B.~Zhang, \href{https://doi.org/10.1093/mnras/stac3599}{Monthly Notices of the Royal Astronomical Society {\bfseries 519}, 1823 (2022)}, arXiv: \href{https://arxiv.org/abs/https://academic.oup.com/mnras/article-pdf/519/2/1823/48447376/stac3599.pdf}{{https://academic.oup.com/mnras/article-pdf/519/2/1823/48447376/stac3599.pdf}}.

\bibitem{chime20repeater}
{The CHIME/FRB Collaboration}, {:}, B.~C. {Andersen}, K.~{Bandura}, M.~{Bhardwaj}, P.~J. {Boyle}, C.~{Brar}, T.~{Cassanelli}, S.~{Chatterjee}, P.~{Chawla}, A.~M. {Cook}, A.~P. {Curtin}, M.~{Dobbs}, F.~A. {Dong}, J.~T. {Faber}, M.~{Fandino}, E.~{Fonseca}, B.~M. {Gaensler}, U.~{Giri}, A.~{Herrera-Martin}, A.~S. {Hill}, A.~{Ibik}, A.~{Josephy}, J.~F. {Kaczmarek}, Z.~{Kader}, V.~{Kaspi}, T.~L. {Landecker}, A.~E. {Lanman}, M.~{Lazda}, C.~{Leung}, H.-H. {Lin}, K.~W. {Masui}, R.~{Mckinven}, J.~{Mena-Parra}, B.~W. {Meyers}, D.~{Michilli}, C.~{Ng}, A.~{Pandhi}, A.~B. {Pearlman}, U.-L. {Pen}, E.~{Petroff}, Z.~{Pleunis}, M.~{Rafiei-Ravandi}, M.~{Rahman}, S.~M. {Ransom}, A.~{Renard}, K.~R. {Sand}, P.~{Sanghavi}, P.~{Scholz}, V.~{Shah}, K.~{Shin}, S.~{Siegel}, K.~{Smith}, I.~{Stairs}, J.~{Su}, S.~P. {Tendulkar}, K.~{Vanderlinde}, H.~{Wang}, D.~{Wulf}, and A.~{Zwaniga}, \href{https://doi.org/10.48550/arXiv.2301.08762}{arXiv e-prints arXiv:2301.08762 (2023)}, arXiv: \href{https://arxiv.org/abs/2301.08762}{{2301.08762}}.

\bibitem{2015arXiv150306462G}
S.~{Gopal Krishna Patro} and K.~K. {Sahu}, \href{https://doi.org/10.48550/arXiv.1503.06462}{arXiv e-prints arXiv:1503.06462 (2015)}, arXiv: \href{https://arxiv.org/abs/1503.06462}{{1503.06462}}.

\bibitem{chawla2002SMOTESyntheticMinority}
N.~V. Chawla, K.~W. Bowyer, L.~O. Hall, and W.~P. Kegelmeyer, \href{https://doi.org/10.1613/jair.953}{Journal of Artificial Intelligence Research {\bfseries 16}, 321 (2002)}.

\bibitem{lemaitre2017ImbalancedlearnPythonToolbox}
G.~Lema{\^i}tre, F.~Nogueira, and C.~K. Aridas, {The Journal of Machine Learning Research {\bfseries 18}, 559 (2017)}.

\bibitem{NIPS2017_7062}
S.~M. Lundberg and S.-I. Lee, {A Unified Approach to Interpreting Model Predictions},  in \emph{Advances in Neural Information Processing Systems 30}, I.~Guyon, U.~V. Luxburg, S.~Bengio, H.~Wallach, R.~Fergus, S.~Vishwanathan, and R.~Garnett, eds., pp.~4765--4774, Curran Associates, Inc., (2017), \href{http://papers.nips.cc/paper/7062-a-unified-approach-to-interpreting-model-predictions.pdf}{http://papers.nips.cc/paper/7062-a-unified-approach-to-interpreting-model-predictions.pdf}.

\bibitem{sun2025exploring}
W.-P. Sun, J.-G. Zhang, Y.~Li, W.-T. Hou, F.-W. Zhang, J.-F. Zhang, and X.~Zhang, {The Astrophysical Journal {\bfseries 980}, 185 (2025)}.

\bibitem{2021ApJ...923....1P}
Z.~{Pleunis}, D.~C. {Good}, V.~M. {Kaspi}, R.~{Mckinven}, S.~M. {Ransom}, P.~{Scholz}, K.~{Bandura}, M.~{Bhardwaj}, P.~J. {Boyle}, C.~{Brar}, T.~{Cassanelli}, P.~{Chawla}, F.~{(Adam) Dong}, E.~{Fonseca}, B.~M. {Gaensler}, A.~{Josephy}, J.~F. {Kaczmarek}, C.~{Leung}, H.-H. {Lin}, K.~W. {Masui}, J.~{Mena-Parra}, D.~{Michilli}, C.~{Ng}, C.~{Patel}, M.~{Rafiei-Ravandi}, M.~{Rahman}, P.~{Sanghavi}, K.~{Shin}, K.~M. {Smith}, I.~H. {Stairs}, and S.~P. {Tendulkar}, \href{https://doi.org/10.3847/1538-4357/ac33ac}{\apj {\bfseries 923}, 1 (2021)}, arXiv: \href{https://arxiv.org/abs/2106.04356}{{2106.04356}}.

\bibitem{hessels2019frb}
J.~Hessels, L.~Spitler, A.~Seymour, J.~Cordes, D.~Michilli, R.~Lynch, K.~Gourdji, A.~Archibald, C.~Bassa, G.~Bower, et~al., {The Astrophysical Journal Letters {\bfseries 876}, L23 (2019)}.

\bibitem{zhou2022fast}
D.~Zhou, J.~Han, B.~Zhang, K.~Lee, W.~Zhu, D.~Li, W.~Jing, W.-Y. Wang, Y.~Zhang, J.~Jiang, et~al., {Research in Astronomy and Astrophysics {\bfseries 22}, 124001 (2022)}.

\bibitem{sasaki2007truth}
Y.~Sasaki et~al., {Teach tutor mater {\bfseries 1}, 1 (2007)}.

\bibitem{PhySO_RL_DA}
W.~{Tenachi}, R.~{Ibata}, and F.~I. {Diakogiannis}, \href{https://doi.org/10.3847/1538-4357/ad014c}{ApJ {\bfseries 959}, 99 (2023)}, arXiv: \href{https://arxiv.org/abs/2303.03192}{{2303.03192}}.

\bibitem{bertrand1878homogeneite}
J.~Bertrand, {Cahiers de recherche de l'Academie de Sciences {\bfseries 86}, 916 (1878)}.

\bibitem{scikit-learn}
F.~Pedregosa, G.~Varoquaux, A.~Gramfort, V.~Michel, B.~Thirion, O.~Grisel, M.~Blondel, P.~Prettenhofer, R.~Weiss, V.~Dubourg, J.~Vanderplas, A.~Passos, D.~Cournapeau, M.~Brucher, M.~Perrot, and E.~Duchesnay, {Journal of Machine Learning Research {\bfseries 12}, 2825 (2011)}.

\bibitem{2022luo}
J.-W. Luo, J.-M. Zhu-Ge, and B.~Zhang, \href{https://doi.org/10.1093/mnras/stac3206}{Monthly Notices of the Royal Astronomical Society {\bfseries 518}, 1629 (2022)}, arXiv: \href{https://arxiv.org/abs/https://academic.oup.com/mnras/article-pdf/518/2/1629/47181506/stac3206.pdf}{{https://academic.oup.com/mnras/article-pdf/518/2/1629/47181506/stac3206.pdf}}.

\bibitem{Chen2022UMAP}
B.~H. {Chen}, T.~{Hashimoto}, T.~{Goto}, S.~J. {Kim}, D.~J.~D. {Santos}, A.~Y.~L. {On}, T.-Y. {Lu}, and T.~Y.~Y. {Hsiao}, \href{https://doi.org/10.1093/mnras/stab2994}{\mnras {\bfseries 509}, 1227 (2022)}, arXiv: \href{https://arxiv.org/abs/2110.09440}{{2110.09440}}.

\bibitem{yang_photo2023}
X.~Yang, S.-B. Zhang, J.-S. Wang, and X.-F. Wu, \href{https://doi.org/10.1093/mnras/stad1304}{Monthly Notices of the Royal Astronomical Society {\bfseries 522}, 4342 (2023)}, arXiv: \href{https://arxiv.org/abs/https://academic.oup.com/mnras/article-pdf/522/3/4342/50258914/stad1304.pdf}{{https://academic.oup.com/mnras/article-pdf/522/3/4342/50258914/stad1304.pdf}}.

\end{thebibliography}\endgroup

\appendix

\section{Robustness to Label Noise}
\label{sec:Robustness}
Repeating bursts reported in Catalog 2 include six special samples that had previously been classified as non-repeating bursts in Catalog 1. Since non-repeating burst data may be subject to observational biases, these samples provide an ideal probe for testing the robustness of our classification model. The NDR equation we developed successfully identified five out of six repeaters, as shown in Table~\ref{tab:six_data}, achieving a score among the best compared with existing machine learning approaches \cite{Chen2022UMAP,2022luo,2022zhuge_nosup,yang_photo2023,sun2025exploring}. Notably, FRB20180910A was consistently classified as a non-repeater across all methodologies examined in this comparative analysis, including our own.

\begin{table}[H]
    \centering
    \renewcommand{\arraystretch}{1.4}  
    \caption{Predictions by Eq. \ref{eq:ndr2} for the six bursts originated classified as non-repeaters in Catalog 1, then reclassified as repeaters in Catalog 2.}
    \label{tab:six_data}
    \begin{tabular}{ @{} l c c @{}} 
        \toprule
        \textbf{FRB Source} & \textbf{Probability} & \textbf{Class} \\
        \midrule
        FRB20180910A & 0.000 & Non-repeater \\
        FRB20190110C & 1.000 & Repeater \\
        FRB20190113A & 0.729 & Repeater \\
        FRB20190226B & 1.000 & Repeater \\
        FRB20190430C & 1.000 & Repeater \\
        FRB20190609C & 1.000 & Repeater \\
        \bottomrule
    \end{tabular}
\end{table}

\section{Repeater Candidates}
\label{sec:Candidates}

We predict potential repeater candidates using the NDR equation. To take advantage of existing studies and provide a cross-check for our predictions, we compare the NDR-predicted list with candidate lists reported in \cite{2022luo} and \cite{2022zhuge_nosup}. Specifically, we construct a cross-validated list by taking the intersection of all three. The NDR equation predicts 85 candidates. The method in \cite{2022zhuge_nosup} predicts 65 candidates, with 30 overlapping with the NDR results. The model in \cite{2022luo} identifies 27 candidates, with 21 overlapping with the NDR results. The three-way intersection yields 17 candidates, which can be considered as high-confidence candidates. Detailed classifications are provided in Table~\ref{tab:frb_candidates}.
\end{multicols}
\begin{longtable}{>{\centering\arraybackslash}p{2.5cm} >{\centering\arraybackslash}p{1cm} >{\centering\arraybackslash}p{1cm} >{\centering\arraybackslash}p{1.5cm} >{\centering\arraybackslash}p{1.5cm} >{\centering\arraybackslash}p{1.0cm} >{\centering\arraybackslash}p{1.5cm} >{\centering\arraybackslash}p{1.8cm} >{\centering\arraybackslash}p{1.5cm}}
\caption{\textbf{Repeater candidates} from the NDR equation.
\checkmark~indicates overlap with \cite{2022luo}, and + with \cite{2022zhuge_nosup}. 
Candidates selected by all three methods are underlined.}
\label{tab:frb_candidates}\\

\toprule
\textbf{Name} & \textbf{$\alpha$} & \textbf{$DM$} & \textbf{$\Delta \nu$} & \textbf{$\Delta t$} & \textbf{$f$} & \textbf{$\nu_p$} & \textbf{Probability} & \textbf{Match} \\
\midrule
\endfirsthead

\toprule
\textbf{Name} & \textbf{$\alpha$} & \textbf{$DM$} & \textbf{$\Delta \nu$} & \textbf{$\Delta t$} & \textbf{$f$} & \textbf{$\nu_p$} & \textbf{Probability} & \textbf{Match} \\
\midrule
\endhead

\midrule
\multicolumn{9}{r}{Continued on next page} \\
\endfoot

\bottomrule
\endlastfoot
    FRB20180725A & 38.2 & 635.4 & 274.8 & 0.000296 & 1.7 & 607.4 & 0.74 &  \\
    FRB20180801A & 60.0 & 547.7 & 209.1 & 0.00058 & 1.11 & 595.6 & 1.00 &  \\
    FRB20180916C & 47.0 & 2166.8 & 277.70 & 0.00406 & 0.39 & 640.9 & 1.00 &  \\
    FRB20180920A & 20.1 & 324.9 & 351.5 & 0.00222 & 0.86 & 585.9 & 0.82 &  \\
    FRB20180920B & 12.3 & 438.3 & 83.20 & 0.00233 & 0.35 & 421.1 & 1.00 & + \\
    FRB20180925A & -1.2 & 159.0 & 400.00 & 0.00452 & 0.99 & 800.2 & 0.99 &  \\
    FRB20181012B & 20.9 & 689.4 & 83.70 & 0.00056 & 0.49 & 428.3 & 0.62 &  \\
    \uline{FRB20181017B} & 61.0 & 269.9 & 205.50 & 0.00231 & 1.06 & 593.2 & 1.00 & $\checkmark$+ \\
    FRB20181018A & 0.6 & 999.7 & 400.00 & 0.0085 & 0.49 & 800.2 & 1.00 &  \\
    FRB20181101A & 16.4 & 1311.1 & 236.6 & 0.00603 & 0.5 & 497.4 & 1.00 &  \\
    FRB20181115A & 19.6 & 951.8 & 168.2 & 0.00183 & 0.44 & 468.8 & 0.99 &  \\
    FRB20181117C & 48.6 & 1705.2 & 259.0 & 0.0001 & 1.57 & 676.5 & 0.99 &  \\
    FRB20181128C & 27.4 & 576.6 & 168.90 & 0.0023 & 0.39 & 480.3 & 1.00 & \checkmark \\
    FRB20181129B & 73.8 & 343.8 & 160.40 & 0.000364 & 4.0 & 556.8 & 1.00 &  \\
    FRB20181203B & 47.8 & 321.5 & 249.40 & 0.000578 & 1.45 & 693.2 & 1.00 &  \\
    FRB20181213B & 45.6 & 604.7 & 270.6 & 0.00085 & 0.75 & 664.0 & 1.00 &  \\
    FRB20181214A & 23.3 & 206.0 & 94.40 & 0.000533 & 0.156 & 435.0 & 0.62 & + \\
    FRB20181214F & 8.0 & 2073.2 & 113.90 & 0.0023 & 0.31 & 425.7 & 0.83 &  \\
    FRB20181218A & 19.0 & 1646.0 & 129.10 & 0.00139 & 0.83 & 448.4 & 0.98 &  \\
    \uline{FRB20181221A} & 62.1 & 292.0 & 137.20 & 0.000754 & 1.25 & 510.1 & 1.00 & \checkmark+ \\
    FRB20181221B & 25.3 & 1333.1 & 192.20 & 0.001037 & 0.97 & 492.2 & 0.91 &  \\
    FRB20181222D & 8.4 & 1394.5 & 161.10 & 0.00375 & 0.22 & 443.0 & 0.99 &  \\
    FRB20181223B & 33.3 & 545.7 & 287.40 & 0.00157 & 0.68 & 600.6 & 1.00 &  \\
    FRB20181228B & 59.3 & 535.2 & 70.30 & 0.0001 & 0.4 & 435.2 & 1.00 &  \\
    \uline{FRB20181229B} & 22.0 & 361.3 & 117.3 & 0.00336 & 0.42 & 445.5 & 1.00 & \checkmark+ \\
    FRB20181230A & 33.0 & 696.6 & 256.70 & 0.00164 & 0.94 & 724.8 & 1.00 &  \\
    \uline{FRB20181231B} & 59.6 & 157.9 & 259.4 & 0.000337 & 0.89 & 657.7 & 1.00 & \checkmark+ \\
    FRB20190101B & 41.7 & 1161.6 & 245.90 & 0.00032 & 1.02 & 713.6 & 0.99 &  \\
    FRB20190102A & 28.9 & 664.2 & 183.6 & 0.000824 & 1.12 & 495.2 & 0.97 &  \\
    FRB20190105A & -1.9 & 343.0 & 400.00 & 0.0039 & 0.6 & 800.2 & 0.70 &  \\
    FRB20190107A & -4.3 & 817.3 & 266.7 & 0.0259 & 0.49 & 400.2 & 1.00 &  \\
    FRB20190110C & 24.5 & 193.0 & 77.5 & 0.00039 & 0.64 & 427.4 & 0.83 & + \\
    \uline{FRB20190112A} & 57.1 & 390.1 & 235.6 & 0.00164 & 1.4 & 697.7 & 1.00 & \checkmark+ \\
    FRB20190113A & 7.3 & 176.4 & 308.30 & 0.00182 & 1.3 & 800.2 & 0.72 &  \\
    FRB20190118B & 13.8 & 627.3 & 367.6 & 0.0037 & 0.31 & 645.0 & 0.99 &  \\
    FRB20190122A & 0.8 & 1175.4 & 400.00 & 0.0055 & 0.327 & 694.2 & 0.97 &  \\
    FRB20190124E & 9.1 & 313.3 & 400.00 & 0.00574 & 0.64 & 625.6 & 0.99 &  \\
    FRB20190125A & 36.0 & 506.4 & 290.1 & 0.00321 & 0.37 & 655.5 & 1.00 & \checkmark \\
    FRB20190128C & 22.6 & 230.4 & 202.6 & 0.00616 & 0.71 & 491.6 & 1.00 & \checkmark \\
    \uline{FRB20190129A} & 43.0 & 437.8 & 247.40 & 0.00113 & 0.49 & 707.7 & 1.00 & \checkmark+ \\
    FRB20190130B & 55.4 & 968.0 & 125.0 & 0.000265 & 0.77 & 487.1 & 1.00 &  \\
    FRB20190131B & 16.5 & 1778.2 & 285.80 & 0.00092 & 0.99 & 798.1 & 0.53 &  \\
    FRB20190206B & 11.6 & 261.0 & 287.40 & 0.0071 & 0.95 & 506.4 & 0.99 & \checkmark \\
    \uline{FRB20190206A} & 38.0 & 152.8 & 201.40 & 0.000804 & 1.4 & 534.5 & 0.99 & \checkmark+ \\
    FRB20190211A & 38.9 & 1050.5 & 285.20 & 0.00036 & 1.47 & 656.0 & 0.99 &  \\
    FRB20190214A & 0.1 & 421.2 & 400.00 & 0.0145 & 0.46 & 800.2 & 1.00 &  \\
    \uline{FRB20190218B} & 46.2 & 482.5 & 231.80 & 0.00205 & 0.57 & 588.0 & 1.00 & \checkmark+ \\
    FRB20190223A & 21.8 & 332.5 & 116.3 & 0.000763 & 0.47 & 444.8 & 0.55 & + \\
    FRB20190226B & 29.9 & 586.2 & 285.9 & 0.004 & 0.38 & 585.2 & 1.00 &  \\
    FRB20190228A & 52.6 & 399.5 & 261.80 & 0.00225 & 1.79 & 664.7 & 1.00 & + \\
    FRB20190304C & 22.3 & 542.5 & 135.00 & 0.000948 & 0.53 & 454.9 & 0.83 &  \\
    FRB20190320D & 9.2 & 1103.9 & 400.00 & 0.0058 & 0.49 & 634.8 & 0.99 &  \\
    \uline{FRB20190329A} & 42.0 & 80.2 & 73.70 & 0.00104 & 0.52 & 432.3 & 1.00 & \checkmark+ \\
    FRB20190403G & 35.7 & 620.0 & 177.70 & 0.00159 & 0.75 & 506.6 & 0.99 &  \\
    FRB20190403E & 31.7 & 181.1 & 316.6 & 0.0022 & 3.9 & 620.6 & 0.99 & + \\
    FRB20190408A & 35.7 & 823.1 & 252.4 & 0.000839 & 0.64 & 576.6 & 0.99 &  \\
    \uline{FRB20190409B} & 21.1 & 244.5 & 286.70 & 0.00234 & 0.39 & 545.5 & 0.99 & \checkmark+ \\
    \uline{FRB20190410A} & 43.0 & 167.5 & 170.5 & 0.00101 & 1.59 & 515.7 & 1.00 & \checkmark+ \\
    FRB20190422A & 42.0 & 367.0 & 279.7 & 0.00322 & 0.6 & 626.1 & 1.00 & + \\
    \uline{FRB20190422A} & 54.2 & 367.0 & 234.2 & 0.00231 & 0.6 & 612.3 & 1.00 & \checkmark+ \\
    FRB20190422A & 24.0 & 367.0 & 318.90 & 0.002 & 0.6 & 582.8 & 0.99 &  \\
    \uline{FRB20190423B} & 62.4 & 160.8 & 159.3 & 0.00249 & 0.87 & 537.6 & 1.00 & \checkmark+ \\
    \uline{FRB20190423B} & 63.0 & 160.8 & 148.5 & 0.0085 & 0.87 & 524.6 & 1.00 & \checkmark+ \\
    FRB20190425B & 22.4 & 982.9 & 172.40 & 0.001108 & 1.25 & 474.8 & 0.73 &  \\
    FRB20190428A & 54.0 & 949.4 & 240.1 & 0.000374 & 2.22 & 696.0 & 0.99 &  \\
    \uline{FRB20190429B} & 99.0 & 263.1 & 42.40 & 0.00638 & 0.74 & 422.4 & 1.00 & \checkmark+ \\
    FRB20190430C & 48.7 & 317.4 & 269.6 & 0.000893 & 2.17 & 659.3 & 0.99 &  \\
    FRB20190519F & 21.7 & 763.2 & 134.2 & 0.00136 & 0.75 & 453.9 & 0.99 &  \\
    FRB20190519J & 24.3 & 589.1 & 60.80 & 0.00046 & 0.63 & 419.5 & 0.99 &  \\
    \uline{FRB20190527A} & 47.0 & 559.9 & 133.70 & 0.00267 & 0.47 & 484.7 & 1.00 & \checkmark+ \\
    FRB20190527A & 30.7 & 559.9 & 112.00 & 0.00247 & 0.47 & 449.1 & 0.99 & + \\
    FRB20190529A & 24.2 & 482.8 & 128.7 & 0.00104 & 0.47 & 453.4 & 0.99 &  \\
    FRB20190531C & 18.3 & 359.4 & 140.40 & 0.00145 & 0.37 & 453.0 & 0.94 & + \\
    FRB20190601B & 9.7 & 754.7 & 115.7 & 0.00404 & 1.0 & 429.5 & 1.00 & + \\
    FRB20190601C & 35.3 & 137.3 & 190.20 & 0.000684 & 1.32 & 517.0 & 0.99 & + \\
    FRB20190601C & 36.1 & 137.3 & 171.80 & 0.00051 & 1.32 & 502.2 & 0.99 & + \\
    FRB20190605D & 48.3 & 1615.9 & 276.20 & 0.001069 & 0.82 & 643.9 & 0.99 &  \\
    FRB20190609A & 62.4 & 259.5 & 192.4 & 0.000432 & 3.6 & 579.3 & 1.00 &  \\
    \uline{FRB20190609A} & 55.0 & 259.5 & 223.30 & 0.00212 & 3.6 & 600.5 & 1.00 & \checkmark+ \\
    FRB20190609C & 15.2 & 328.0 & 81.10 & 0.00207 & 0.64 & 422.9 & 0.99 &  \\

    FRB20190617B & 10.6 & 225.0 & 186.50 & 0.00758 & 0.99 & 459.3 & 0.99 & + \\
    FRB20190617C & 10.4 & 602.9 & 363.30 & 0.00388 & 0.54 & 732.3 & 0.99 &  \\
    FRB20190621C & 39.1 & 547.8 & 146.90 & 0.000443 & 1.98 & 485.4 & 0.99 &  \\
    FRB20190623B & 54.2 & 1422.0 & 260.20 & 0.00044 & 1.58 & 643.3 & 0.99 &  \\
    FRB20190701C & 46.2 & 916.6 & 93.3 & 0.00144 & 0.88 & 446.4 & 1.00 &  \\
\end{longtable}

\begin{multicols}{2}

\end{multicols}

\end{document}